\documentclass[12pt]{iopart}

\usepackage{booktabs}
\usepackage{colortbl}  % For coloring table cells
\usepackage{xcolor}    % For defining colors
\usepackage{caption}
\usepackage{graphicx}

\definecolor{lightgray}{gray}{0.9}
\definecolor{lightblue}{rgb}{0.8, 0.85, 1}

\begin{document}

\title[]{Integrated Modeling of SPARC H-mode Scenarios: Exploration of the Impact of Modeling Assumptions on Predicted Performance}

\author{M. Muraca$^1$, P. Rodriguez-Fernandez$^1$, N.T. Howard$^1$, J. Hall$^1$, E. Fable$^2$, G. Tardini$^2$}

\address{$^1$MIT Plasma Science and Fusion Center, 167 Albany St, Cambridge, MA 02139, USA \\
$^2$Max Planck Institut fuer Plasma Physik, Boltzmannstrasse 2, Garching bei Muenchen, 85748, Germany}
\ead{mmuraca@mit.edu}
\vspace{10pt}
\begin{indented}
\item[]November 2024
\end{indented}

\begin{abstract}
In this paper an extensive database of SPARC H-modes confinement predictions has been provided, to assess its variability with respect to few input assumptions. The simulations have been performed within the ASTRA framework, using the quasi-linear model TGLF SAT2, including electromagnetic effects, for the core transport, and a neural network trained on EPED simulations to predict the pedestal height and width self-consistently. The database has been developed starting from two SPARC H-mode discharges (12.2 T, i.e. Primary Reference Discharge or PRD, and 8 T, i.e. reduced field) and permuting 4 input parameters (W concentration, DT mixture concentration, temperature ratio at top of pedestal and deviation of pedestal pressure from the EPED prediction), to perform a sensitivity study. For the PRD a scan of auxiliary input power (ion cyclotron heating) has been performed up to 25MW, to keep highly radiative plasmas above the LH power threshold as predicted by Martin and Schmidtmayr power scalings. A scan of pedestal density has then been performed for both PRD and 8T databases. $p_{top}/p_{EPED}$ and $T_i/T_e$ at top of pedestal showed the biggest impact on the fusion gain. Significant variation is observed across the database, highlighting the importance of sensitivity studies. Below a certain W concentration, the 12T database shows that $Q>5$ is consistently achieved for full-field H-modes with 11 MW of auxiliary power, and values of $Q>2$ are assured when increasing the input power to keep the plasma in H-mode. The 8T database demonstrates that SPARC can access a $Q>1$ operational window with low W concentration, making it a potentially interesting scenario for obtaining breakeven conditions.
\end{abstract}

%\maketitle

\section{Introduction}
The prediction of plasma fusion performance is crucial for both current experiments and future devices. Beyond revealing fascinating physics, it provides a key benchmark for assessing research progress and validating the reliability of modeling approaches. Fusion performance is predominantly constrained by the level of transport, which arises from a complicated interplay between mechanisms that drive and suppress turbulence. Various approaches exist for modeling transport, differing in the level of detail included in their descriptions and the computational resources required. In this work, we define \textit{high-fidelity} simulations as those capable of describing the physics of a process with great detail (e.g. kinetic and gyrokinetic nonlinear models of electromagnetic plasma turbulence) \cite{jenko_massively_2000, peeters_effect_2004, kotschenreuther_comparison_1995, candy_eulerian_2003, grandgirard_drift-kinetic_2006}. These models, while accurate, are often impractical due to their significant computational cost. High-fidelity simulations can predict transport under specific assumptions and, when integrated within a framework that accounts for the evolution of power sources and sinks, can be used to evaluate plasma performance \cite{candy_tokamak_2009, barnes_direct_2010, rodriguez-fernandez_enhancing_2024}. On the other hand, \textit{Low-fidelity} simulations typically rely on scaling laws, experimental trends and simplified models, which derive global 0-D quantities starting from engineering and control parameters \cite{houlberg_contour_1982, slendebroek_elevating_2023}. Examples of such an approach are Plasma OPerational CONtour (POPCON) diagrams \cite{houlberg_contour_1982}, which have been used for SPARC predictions \cite{creely_overview_2020}. These are the fastest models, but they are prone to the largest uncertainties, particularly as one extrapolates to reactor-relevant conditions. Low-fidelity simulations are particularly useful in the design phase of a machine or during pulse planning, where they enable a broad exploration of the operational space, or where empirical data is plentiful. This allows for the rapid selection of an operational point that optimizes metrics such as fusion power or gain. \\
Once an operational point is identified, it is essential to acknowledge that performance parameters are derived using simplified models. Consequently, benchmarking these results against more reliable models, such as gyrokinetic simulations, becomes necessary. However, due to the inherent complexity of reproducing all the physics self-consistently, both high and low fidelity simulations require assumptions on certain physics parameters, e.g. impurity concentrations, which can influence the outcomes. Studying the sensitivity of these results to the assumed input parameters is therefore crucial. This task can be efficiently addressed by \textit{medium-fidelity} models, which represent a good compromise between level of detail (i.e. reliability of the results) and time consumption. In fact, medium-fidelity simulations can be run with reduced physics-based models (e.g. quasi-linear transport models), speeding up the simulations by several orders of magnitude \cite{staebler_verification_2021, snyder_first-principles_2011, bourdelle_core_2016}. These models are often employed in frameworks able to describe the evolution of power sinks and sources, enabling the determination of plasma energy and particle confinement \cite{pereverzev_astra_1991, pankin_transp_2024, romanelli_jintrac_2014}. An example of an integrated framework using medium-fidelity models is presented in \cite{luda_validation_2023}. The ability to predict fusion performance based on assumed input parameters and assess its variability is crucial for high-field compact devices like SPARC \cite{creely_overview_2020}, since they operate at high plasma density and fusion power, representing a novelty with respect to previous experiments, and missing therefore a strong experimental support. \\
In this work we investigate the sensitivity of fusion performance parameters for 2 SPARC H-mode scenarios, corresponding to nominal and reduced toroidal magnetic field (12 and 8 T). The results reveal significant variability of the performance, demonstrating that the choice of input parameters has a substantial impact on the predicted fusion gain. \\
The remainder of the paper is organized as follows: in section 2, the framework of the simulations is described; in section 3, the strategy to assess the performance variability with respect to the input assumptions is presented; in section 4, the SPARC Primary Reference Discharge (PRD) is introduced and the uncertainty quantification of its performance is discussed; in section 5, the the reduced field H-mode is described and the results of its performance quantification are shown; in section 6, the article is summarized and the conclusions are discussed, together with an outlook for future work.

\section{Simulation Setup}
In order to evaluate the confinement and performance of a discharge, a collection of models needs to be integrated in a framework capable of predicting plasma kinetic profiles. This process requires the simultaneous evolution of power/particle target fluxes (i.e. sum of local sources and sinks of power and particles) and transport fluxes, until the two converge to the same value. This work employs the time-dependent transport solver ASTRA \cite{pereverzev_astra_1991}, to accomplish this task. Although ASTRA can simulate the time evolution of a plasma pulse, this work focuses on the stationary flat-top phase of two H-mode scenarios, where the fusion power has reached a relaxed state. Due to its modular and flexible nature, ASTRA enables the implementation of custom models, including Neural Networks (NNs). The equilibrium of the plasma in ASTRA has been calculated with SPIDER \cite{ivanov_new_2005}, which performs fixed-boundary calculations based on separatrix coordinates derived from earlier freeGS \cite{noauthor_freegs-plasmafreegs_2024} simulations.\\
Core transport in tokamaks is predominantly associated with gradient-driven micro-instabilities like Electron Temperature Gradient (ETG), Trapped Electron Mode (TEM) and Ion Temperature Gradient (ITG) \cite{weiland_collective_2000, dimits_comparisons_2000, garbet_introduction_2006}. For this study, turbulence and transport in the plasma core are predicted using the quasi-linear TGLF model \cite{staebler_verification_2021}. TGLF has been extensively validated on various machines and plasma scenarios worldwide, establishing its robustness and reliability \cite{rodriguez-fernandez_predict-first_2019, rodriguez-fernandez_perturbative_2019, angioni_confinement_2022, baiocchi_turbulent_2015}. The control settings include electromagnetic (EM) effects, three plasma species, Miller geometry \cite{stacey_representation_2009}, saturation rule SAT2 \cite{staebler_verification_2021} and a maximum of 6 parallel basis functions.\\
The scenarios reproduced in this project are H-modes; therefore, assessing the pedestal stability is essential, according to peeling-ballooning theory. This task is performed using a NN trained on EPED \cite{snyder_first-principles_2011} results for SPARC, following a methodology similar to \cite{meneghini_self-consistent_2017}. The NN model was validated on a sample of over 2000 unpublished EPED simulations with a mean squared error (MSE) of less than 10\%. It is important to note that the ELMy H-mode pedestal predicted by EPED is likely to be an upper limit for safe operation. The model updates self-consistently the pedestal height as the plasma gradients and overall stored energy change. This is crucial because the EPED pedestal height is directly related to the normalized beta, therefore fixed boundary conditions may result in an over- or underprediction. In the NN model the pedestal density is an input, as it depends on fueling, neutral penetration and edge physics, which is not described in this work. \\
SPARC will operate with Ion Cyclotron Resonance Heating (ICRH) \cite{lin_physics_2020}. Due to the absence of a self-consistent model to reproduce the ICRH absorption profiles directly within ASTRA, simulations were performed using TRANSP \cite{pankin_transp_2024}, coupled with TORIC \cite{brambilla_numerical_1999} and FPPMOD \cite{hammett_fast_1986}, for ICRF wave propagation and Fokker Planck collisions calculations, respectively. The resulting ICRH deposition profiles (to main ions and electrons, including the minority contribution) are imported into ASTRA simulations. The other sources and sinks of power are computed using analytical formulae. Specifically, the ohmic power is $P_{OH}=\sigma E_z^2$, where $\sigma$ is the plasma conductivity and $E_z$ is the electric field in the toroidal direction, the collisional exchange is $P_{ex}=0.00246 \Lambda n_e n_i Z^2(T_e-T_i)(A T_e^{3/2})^{-1}$, where $\Gamma$ is the Coloumb logarithm, $n_e$ and $n_i$ ar the electron and ion densities, $T_e$ and $T_i$ are the electron and ion densities, $Z$ and $A$ are the charge and mass of the main ion species, the fusion power is calculated as in \cite{barnett_plasmas_2013}, and $P_{rad}$ is the sum of the synchrotron, Bremmstrahlung and W radiation terms, with the latter calculating following \cite{putterich_calculation_2010}. Due to the uncertainties in the neutral source penetration and ionization models, especially in the stiff pedestal region characteristic of H-modes, the density profile may result only by the balance between inward convection (i.e. pinch) and outward diffusion, matching a null particle flux (source-free condition) in the core region, just inside the pedestal, where a density boundary condition is assumed. Hence, no core particle source has been used in the simulations.\\
SPARC's compact design and high particle and power fluxes necessitate careful management of divertor loads to mitigate damage to Plasma-Facing Components (PFCs). Detachment is one potential strategy to protect the divertor and is often achieved by pumping impurities into the Scrape-Off Layer (SOL). A fraction of these impurities penetrate into the core, reducing performance via radiation. The presence of Tungsten (W) and Boron (B), introduced by erosion of PFCs and boronization, as well as plasma minority ions used for efficient ICRH heating \cite{van_eester_minority_2012}, further influences core performance. Thus, it is important to include impurities in the simulations, to account for their impact on radiation and transport through $Z_{eff}$ and main species dilution. While TGLF can predict impurity transport, estimating impurity sources requires additional modeling beyond the scope of this work. Therefore, the impurities have been included in the simulations by assuming a uniform concentration relative to the electron density profile.\\
Sawteeth are expected to affect the shape of core kinetic profiles in SPARC plasmas, particularly in the absence of neutral beams and current drive \cite{rodriguez-fernandez_predictions_2020}. Although this study does not focus on magnetohydrodynamics (MHD), accurately calculating the safety factor profile is crucial because it directly affects transport. Therefore, the Kadomtsev model \cite{kadomtsev_disruptive_1975} has been used in ASTRA to predict the sawtooth mixing radius and flatten the safety factor and temperature profiles. Additionally, ASTRA requires the sawtooth period as an input to trigger the profile relaxation. This period is found equal to 1 second, using the Porcelli model \cite{porcelli_model_1996} within a TRANSP simulation, and subsequently used as fixed parameter in the ASTRA simulation database. A sensitivity study of the sawtooth period, performed with a limited number of ASTRA runs, reveals a very weak correlation with the resulting fusion power.

\section{Methodology for the Performance Database Generation}
Transport simulations always require input variables such as plasma geometry, auxiliary power, control system settings and other engineering parameters. These quantities are typically selected and controlled to guide the pulse evolution and handle the behaviour of the plasma. In an ideal model that accounts for all physical processes occurring within the tokamak, no additional assumptions would be required, and all parameters would be determined self-consistently. However, current models lack the predictive capability to fully capture all the physical processes that govern specific quantities. As a result, simulations must rely on certain quantities as assumed input parameters, as they are inherently uncertain and challenging to control during the discharge. Hence, it is crucial to examine the impact of the assumed parameters, particularly because they play a significant role in the determination of fusion performance. In this work, we have selected four key plasma parameters, whose values exhibit considerable uncertainty and are hypothesized to significantly influence the predictions of plasma performance:
\begin{itemize}
    \item the W concentration, i.e. $f_W=n_W/n_e$. This parameter mainly affects the radiation, due to the high atomic number of the species. Being present as a trace of the main species, its effect on transport through dilution or $Z_{eff}$ is small;
    \item the DT fraction, i.e. $f_{DT}$. The discharges analyzed in this paper contain 50\% D and 50\% T fuel. $f_{DT}$ is the amount of DT fuel with respect to the electron density, not the specific fuel composition. At constant temperature, fusion power increases proportionally to the DT concentration. However, increasing the DT concentration has been demonstrated to increase turbulent transport \cite{ennever_effects_2015, kim_full-f_2017, rodriguez-fernandez_core_2024};
    \item the ratio of ion to electron temperature at the top of pedestal, i.e. $T_{i,top}/T_{e,top}$. This parameter affects the fusion performance in two key ways: a higher ion temperature drives higher fusion power; higher $T_i/T_e$ in the core can raise the ITG linear threshold \cite{jenko_critical_2001} and reduce its associated transport, which is dominant in SPARC H-mode scenarios \cite{rodriguez-fernandez_overview_2022}. In the EPED framework, $T_e=T_i$ is often assumed, and the pressure at top of pedestal is calculated by assuming an input pedestal density. However, in this work we permute the temperature ratio at top of pedestal, keeping a total pressure consistent with the EPED-NN prediction. This choice is motivated by experimental observations of $T_i\neq T_e$ at the pedestal of current devices \cite{ho_neural_2021}, and can not be excluded for future devices, given their low collisionality. Generally, it is important to note that the $T_i=T_e$ assumption is contingent upon the ratio $\tau_{ex}/\tau_e$, where $\tau_{ex}$ is the collisional time and $\tau_e$ the confinement time;
    \item the ratio between top of pedestal pressure and its value calculated by EPED. This ratio is assumed as input due to the expected error bars of EPED predictions when validated against experiments [cite]. Assuming a variability of the pedestal with respect to the predicted value is particularly important for future devices, for which EPED has not yet been validated. This ratio affects the performance, since a higher pedestal sustains overall higher kinetic profiles, driving more fusion power.
\end{itemize}
The introduced input parameters have standard assumptions considered for both SPARC H-mode scenarios, which are used by previous modeling \cite{creely_overview_2020}. They are listed in table \ref{tab:refPRD}. $f_{DT}$ for the 8T SPARC H-mode, whose value is $0.9$, differs from the SPARC PRD value due to the adoption of H instead of He3 minority species for efficient heating.
\begin{table}[h]
  \centering
  \caption{Standard input assumptions for the SPARC PRD simulation.}
  \begin{tabular}{ccccc}
    \toprule
    \rowcolor{lightblue} % Row color for header
    \textbf{} & \textbf{$f_{DT}$} & \textbf{$f_W$} & \textbf{$\frac{T_i}{T_e}$} & \textbf{$\frac{p_{top}}{p_{eped}}$} \\
    \midrule
    \textbf{12.2T} & 0.85 & 1.5$\cdot 10^{-5}$  & 1 & 1              \\
    \textbf{8T} & 0.9 & 1.5$\cdot 10^{-5}$  & 1 & 1              \\ % Replace '--' with actual data
    \bottomrule
  \end{tabular}
  \label{tab:refPRD}
\end{table}
Given that the exact values of the discussed quantities are difficult to predict, ranges of variations, loosely based on experimental evidence from current experiments, are assumed. The ranges are listed in table \ref{tab:ranges}.
\begin{table}[h]
  \centering
  \caption{Ranges of variation of the input assumptions for SPARC PRD and 8T simulations.}
  \begin{tabular}{ccccc}
    \toprule
    \rowcolor{lightblue} % Row color for header
     & \textbf{$f_{DT}$} & \textbf{$f_W$} & \textbf{$\frac{T_i}{T_e}$} & \textbf{$\frac{p_{top}}{p_{eped}}$} \\
    \midrule
            \textbf{12.2T} & $\left[0.8-0.875\right]$ & $\left[ 1.5\cdot 10^{-5} - 1.35\cdot 10^{-4}\right]$  & $\left[0.8-1.2\right]$ & $\left[0.8-1.2\right]$        \\
            \textbf{8T} & $\left[0.85-0.925\right]$ & $\left[ 1.5\cdot 10^{-5} - 1.35\cdot 10^{-4}\right]$  & $\left[0.8-1.2\right]$ & $\left[0.8-1.2\right]$        \\
    \bottomrule
  \end{tabular}
  \label{tab:ranges}
\end{table}
The species treated in a plasma simulation need to satisfy quasi-neutrality, implying that their concentrations must fulfill a criterion and can not assume any randomly assigned value. In order to overcome this difficulty, a \textit{lumped} impurity has been modeled in this study. This impurity represents an average of all the other species present in the tokamak (typically B, O and C, together with any element used to seed/puff in the SOL). The concentration and atomic number of the lumped impurity can be determined applying two constraints, which are the quasi-neutrality and a specific value of $Z_{eff}$. $Z_{eff}=1.5$ has been selected for the standard input assumptions to reflect a reduced core penetration of light species. While the assumption of $Z_{eff}$ aligns with recent ITER simulations \cite{fajardo_integrated_2024}, a more accurate evaluation requires modeling of the seeding and divertor handling, which is beyond the scope of this paper. The choice of minority species, such as He3 for PRD and H for 8T H-mode, with $f_{minority}=0.05$ \cite{lin_physics_2020}), results in different DT concentrations for the reference values of the input parameters listed in \ref{tab:refPRD}. The lumped impurity atomic number is fixed across the simulations, allowing $Z_{eff}$ to vary within a reasonable range, while the impurity concentration is adjusted to ensure quasi-neutrality for other combinations of $f_{DT}$ and $f_W$. The same atomic number (i.e. 8) is used for the lumped impurity in both the PRD and 8T H-mode scenarios. \\
To assess their effect on the performance, at the beginning of each simulation the uncertain input parameters are randomly assigned (with a uniform distribution) in the ranges discussed, enforcing the described constraints. Based on the setup outlined in the current and previous sections, two databases have been constructed for the SPARC PRD and the 8T H-mode scenarios. Following the initial group of simulations, scans in the ICRH input power and pedestal density were conducted. The results of these scans indicate a favorable operational point where the risk of low performance is minimized, ensuring safe and stable operation of the machine. This scoping process can assist in guiding the experimental campaign toward physics-aware scenario modeling and pulse planning. The results are presented in the following two sections for the two distinct discharges.

\section{SPARC Primary Reference Discharge}
The SPARC PRD scenario has been already explored with low and high fidelity simulations \cite{creely_overview_2020}, \cite{rodriguez-fernandez_overview_2022}. Its main global parameters are listed in table \ref{tab:PRDvalues} and their definition can be found in \cite{creely_overview_2020}.

\begin{table}[h]
  \centering
  \caption{List of the main global and engineering parameters of the SPARC Primary Reference Discharge. Their definition is in \cite{creely_overview_2020}}
  \begin{tabular}{cc}
    \toprule
    \rowcolor{lightblue} % Row color for header
    \textbf{Parameter} & \textbf{Value} \\
    \midrule
          $B_t$  &  12.2 T  \\
          $I_p$  &  8.7 MA  \\
          $R_0$  &  1.85 m  \\
          $a$  &  0.57 m  \\
          $k_{sep}$  &  1.97  \\
          $\delta_{sep}$  &  0.54  \\
          $P_{ICRH}$  &  11 MW  \\
          $P_{ICRH,max}$  &  25 MW  \\
          $\langle n_e \rangle$  &  $3.1\cdot10^{20}m^{-3}$  \\
          $\Delta t_{flattop}$  &  10 s  \\
          $q^*_{Uckan}$  &  3.05  \\
          $f_G$  &  0.37  \\
    \bottomrule
  \end{tabular}
  \label{tab:PRDvalues}
\end{table}

\begin{figure}[h]
    \centering
    \includegraphics[width=0.45\textwidth]{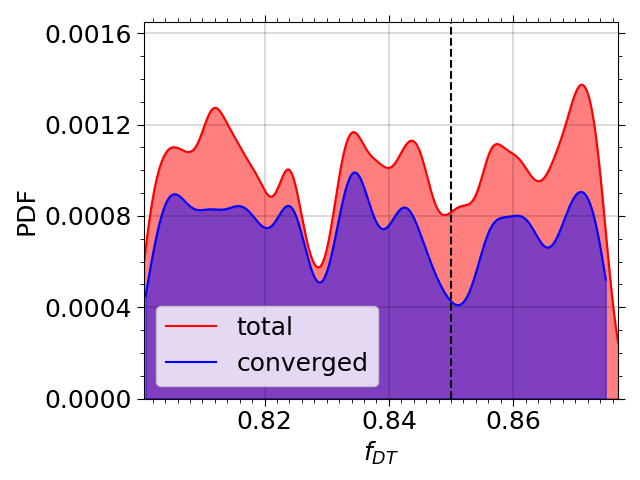}
    \includegraphics[width=0.45\textwidth]{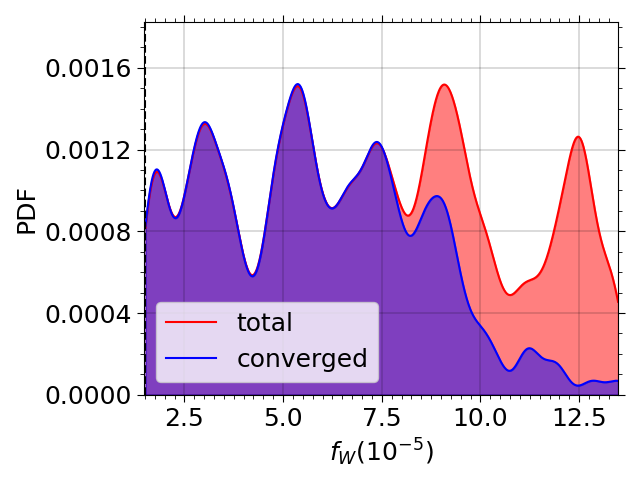}
    \includegraphics[width=0.45\textwidth]{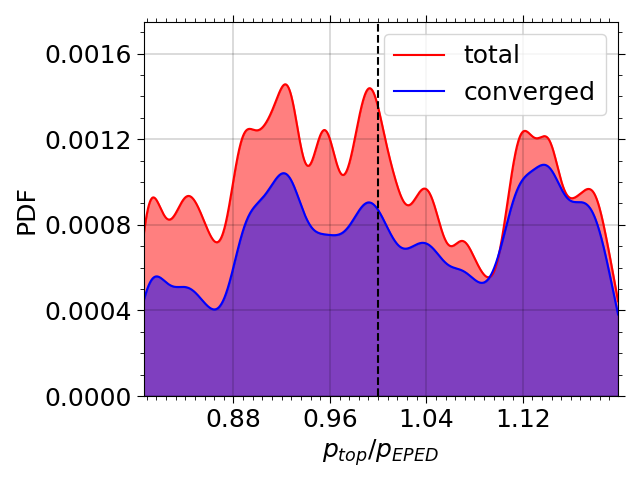}
    \includegraphics[width=0.45\textwidth]{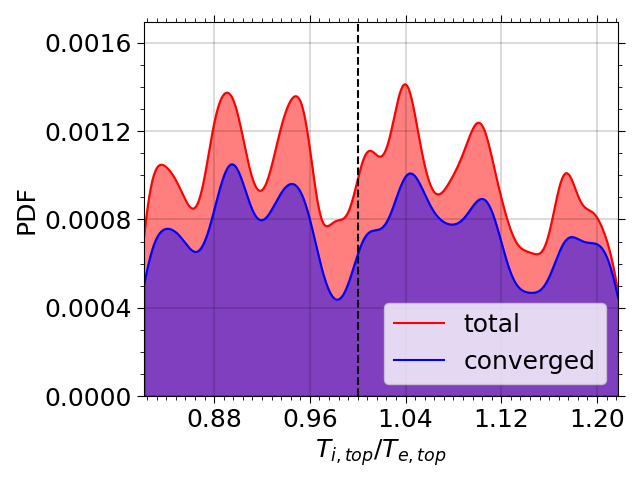}
    \caption{PDF of the 4 input parameters changed within the initial database of 284 simulations, generated with a bandwidth of $0.08$. These parameters are assigned randomly and uniformly. Here, in red (blue) the distribution of inputs for the total launched (only converged) simulations are shown. 74\% of the total simulations converged. One can see the strong asymmetry of the blue curve in the $f_W$ plot, due to the fact that a high W content can radiate strongly, with a consequent plasma collapse. The dashed lines indicate the reference input assumptions.}
    \label{fig:PRD_ranges}
\end{figure}

\begin{figure}[h]
    \centering
    \includegraphics[width=0.45\textwidth]{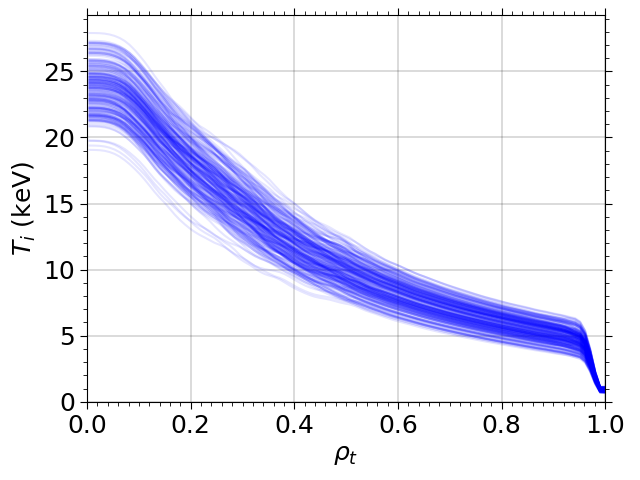}
    \includegraphics[width=0.45\textwidth]{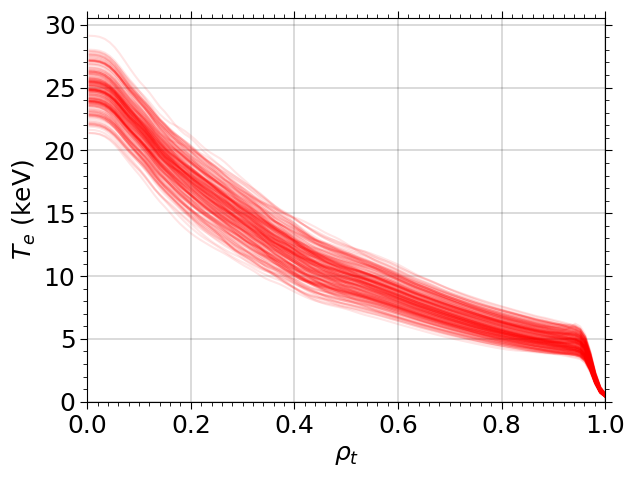}
    \includegraphics[width=0.45\textwidth]{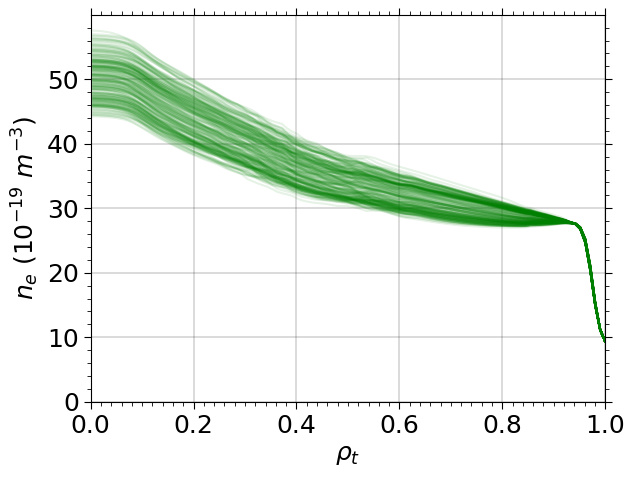}
    \caption{Kinetic profiles of the PRD simulations varying the input assumptions, at constant input power (11MW) and density at top of pedestal ($2.7$ $10^{20}m^{-3}$).}
    \label{fig:PRD_kinetic_profiles}
\end{figure}

\begin{figure}[h]
    \centering
    \includegraphics[width=0.45\textwidth]{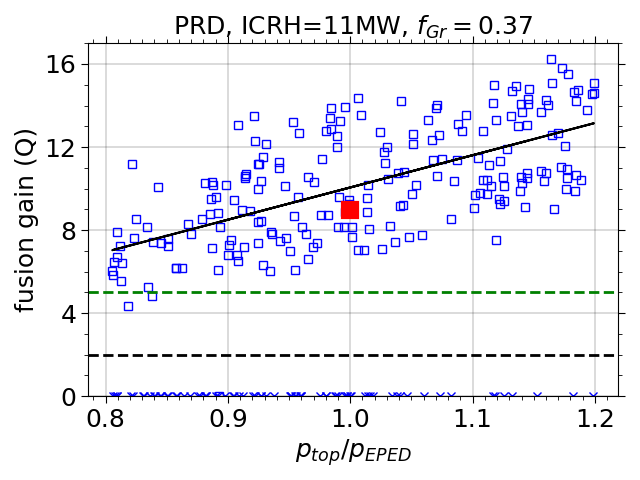}
    \includegraphics[width=0.45\textwidth]{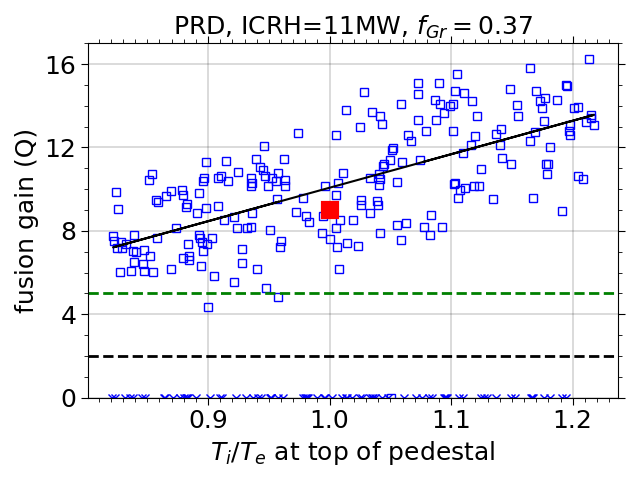}
    \caption{Q vs $p_{top}/p_{EPED}$ and $T_{i,top}/T_{e,top}$ for an initial database of PRD performance with $P_{ICRH}=11MW$ and $f_G=0.37$. The 2 input parameters shown in these plots are the ones which mainly affect the performance. The solid black line indicates the linear trend which fits the scattering of data (ignoring the collapsed points, which are on the horizontal Q=0 line). The black (green) dashed line indicates Q=2 (Q=5). Almost every converged point lay above these values. The red square is the PRD with standard assumptions of the input parameters.}
    \label{fig:Q_trends}
\end{figure}

For this scenario an optimal point has been found scoping the operational space with POPCON \cite{creely_overview_2020}. The primary engineering input parameters are $P_{ICRH}=11MW$ and $f_G=0.37$ and the predicted Q is 11.
%
%\begin{figure}[h]
%    \centering
%    \includegraphics[width=0.6\textwidth]{Screenshot 2024-06-26 at 3.25.30 PM.png}
%    \caption{SPARC PRD operational space, scoped by POPCON. Figure as courtesy from \cite{creely_overview_2020}}
%    \label{fig:popcon}
%\end{figure}
%
This operational point was identified under specific assumptions, which may not be guaranteed or easily controlled during the machine's operation. It is, therefore, important to assess the variability of the performance with respect to the values of these input assumptions. The input parameters were varied uniformly and randomly in the ranges shown in figure \ref{fig:PRD_ranges}, represented through probability distribution functions (PDFs).
In this figure the red curve represents the total distribution of inputs, while the blue curve corresponds to the distribution of inputs from simulations that achieved a converged flux-matched condition. While for most of the parameters the two distribution shapes are similar, $f_W$ shows a strong asymmetry for the blue curve. In other words, at high W concentration there are less converged simulations, due to the associated high W radiation. As will be discussed further, this parameter is crucial in the analysis. At higher pressures there are more converged simulations, due to the higher fusion power, which can compete with higher radiative power levels, suggesting that $p_{top}/p_{EPED}$ plays an important role. Despite the distribution curves not being flat, all input values are represented with a non-negligible statistical weight. However, the available computing resources did not permit additional simulations. Better coverage of the input parameter space is planned for future work to ensure more robust uncertainty quantification.
Using these inputs, an initial database of 284 simulations has been obtained ($\sim 27000$ CPUh). Within the database, 74\% of the simulations converged, while the rest collapsed due to the high W concentration. The kinetic profiles of the converged cases are shown in figure \ref{fig:PRD_kinetic_profiles}.
In figure \ref{fig:Q_trends} the dependence of the predicted Q with respect to $p_{top}/p_{EPED}$ and $T_{i,top}/T_{e,top}$, which are the 2 main parameters affecting the performance, is shown. Here Q is defined as $P_{fus}/(P_{aux}+P_{Ohm})$. Despite significant scatter, a linear increase in performance with increasing pedestal pressure is evident, as expected, due to the overall higher sustained kinetic profiles. Similarly, Q increases with higher $T_{i,top}/T_{e,top}$, due to the enhanced fusion power. One could argue that part of the improvement with a high ion to electron temperature ratio stems from improved ITG stability. However, while the analysis of the TGLF spectra for these simulations revealed strong ITG dominance, the temperature ratio has a weak effect due to the simultaneous strong variability of the other parameters. A detailed analysis of the spectra is presented in section 4.4. One can see that all the converged points show $Q>2$ (black dashed line) and almost all of them have burning plasma condition (i.e. $Q>5$, green dashed line).

\subsection{Effect of W radiation}
\begin{figure}[h]
    \centering
    \includegraphics[width=0.45\textwidth]{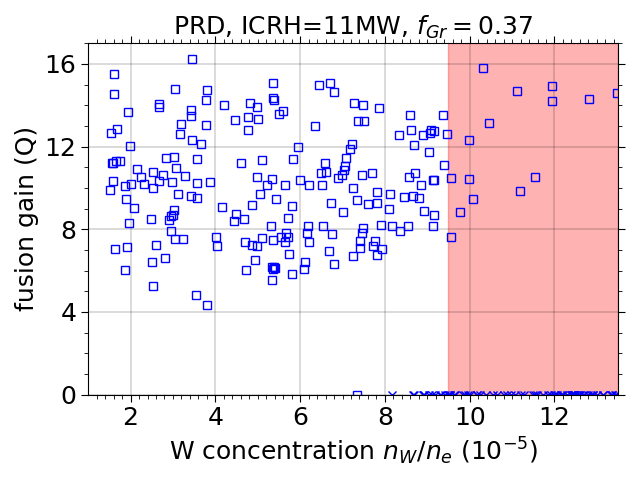}
    \includegraphics[width=0.45\textwidth]{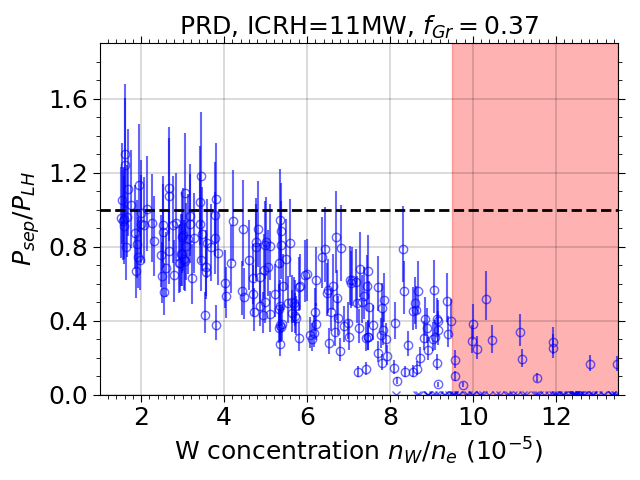}
    \caption{Q vs $f_W$ on the left, $P_{sep}/P_{LH}$ vs $f_W$ (where $P_{LH}$ is calculated with the Martin scaling) on the right, for an initial database of PRD performance with $P_{ICRH}=11MW$ and $f_G=0.37$. In red it is shown the region where the W concentration becomes not tolerable due to the high radiation. The points laying at $Q=0$ show unconverged simulations, where a radiative collapse occurred. The black dashed line on the right plot shows $P_{sep}/P_{LH}=1$ which indicates a sustained H-mode according to the Martin scaling.}
    \label{fig:fW_trends}
\end{figure}

$f_W$ is a crucial input parameter. In fact, this quantity plays a key role in determining the amount of radiation, and if too high it can lead to an intolerable power sink and a radiative collapse, resulting in a unconverged simulation. Moreover, the power at the separatrix ($P_{sep}$) is impacted by the radiation, which is particularly important for H-mode simulations. For current machines, statistical analysis of experimental data have identified $P_{sep}$ thresholds (i.e. $P_{LH}$), above which the plasma can access H-mode. Examples of these scalings include Martin \cite{martin_power_2008} and Schmidtmayr \cite{schmidtmayr_investigation_2018} formulae, which differ primarily in their treatment of total versus ion power at the separatrix, respectively. Additionally, the Martin scaling is based on data from carbon-wall machines, while the Schmidtmayr scaling incorporates data from both carbon- and metal-wall devices. In the following, the ratio between $P_{sep}$ and $P_{LH}$ will be referred to as $f_{LH}$.\\
In figure \ref{fig:fW_trends} Q vs $f_W$, and $P_{sep}/P_{LH}$ vs $f_W$ are shown. On the left plot, there is no clear trend of the fusion gain with respect to the W concentration. This lack of correlation arises partly because other input parameters have a more significant impact on overall performance. Additionally, the flat trend can be attributed to profile stiffness, a well-known phenomenon observed experimentally \cite{imbeaux_modelling_2001}, \cite{ryter_electron_2003}. The stiffness implies that as long as the micro-instabilities governing the transport in the core of a plasma are above their linear thresholds, the resulting kinetic profiles are nearly independent of the injected power source. Importantly, figure \ref{fig:fW_trends} shows that the number of converged points increases at lower $f_W$ values compared to higher $f_W$. This occurs because, beyond a certain $f_W$, most simulated conditions experience radiative collapse, resulting in unconverged simulations (points with $Q=0$ in the left plot). A weak relation between $f_W$ and the plasma performance was also found for DEMO \cite{fable_selected_2017}. The right plot indicates that as $f_W$ increases, the plasma fails to sustain a robust H-mode according to the Martin scaling (i.e. $P_{sep}<P_{LH}$). Calculating the threshold with the Schmidtmayr scaling leads to the same trend.

\subsection{ICRH Power Scan}
\begin{figure}[h]
    \centering
    \includegraphics[width=0.45\textwidth]{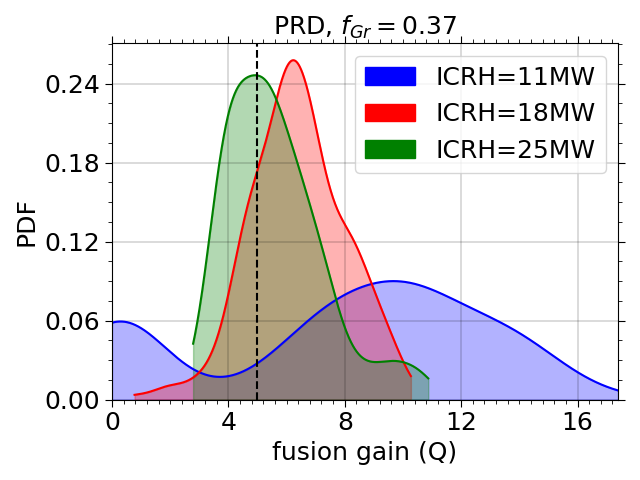}
    \includegraphics[width=0.45\textwidth]{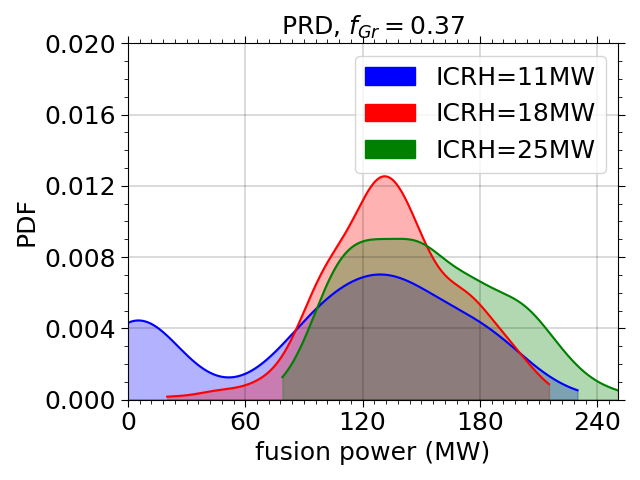}
    \includegraphics[width=0.45\textwidth]{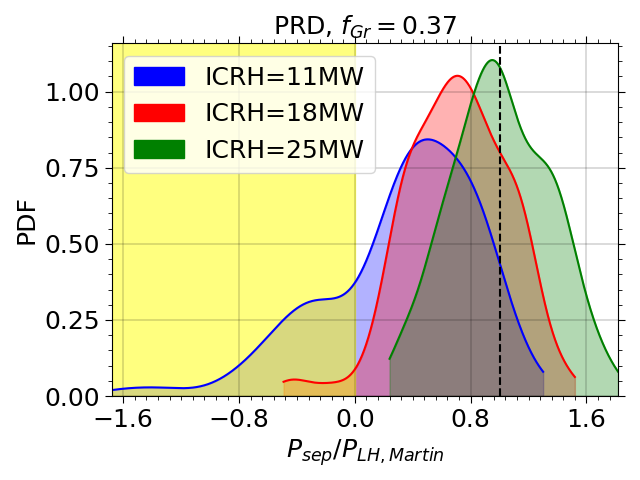}
    \includegraphics[width=0.45\textwidth]{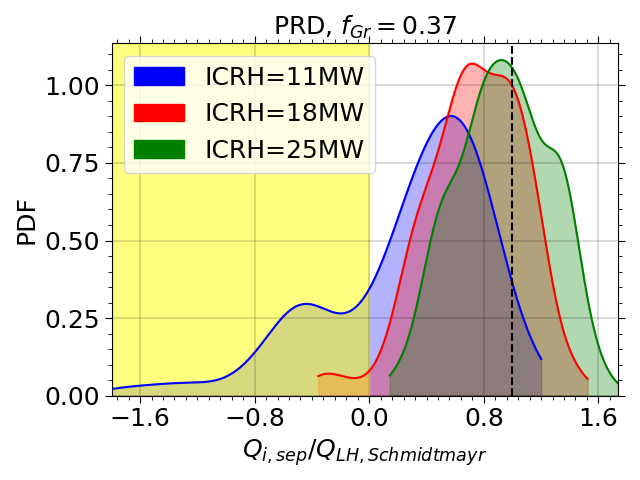}
    \caption{Probability distribution function of different parameters for an input ICRH power scan. Starting from the upper left, clockwise: fusion gain, fusion power, $f_{LH}$ using an ion heat flux threshold, $f_{LH}$ using a total flux threshold. The yellow regions of the plots indicate unconverged simulations. One can see that increasing the power does not change the overall fusion power because of the stiffness of the kinetic profiles, but the fusion gain is reduced. However, $Q>3$ is found for all the converged simulations, whose number increases with the input power. The bottom plots show that with increased power the H-mode is better sustained. Starting from the upper left, clockwise, the dashed black vertical line indicates: burning plasma condition and sustained H-mode condition with total and ion power scaling.}
    \label{fig:PRD_PDFs}
\end{figure}

\begin{figure}[h]
    \centering
    \includegraphics[width=0.45\textwidth]{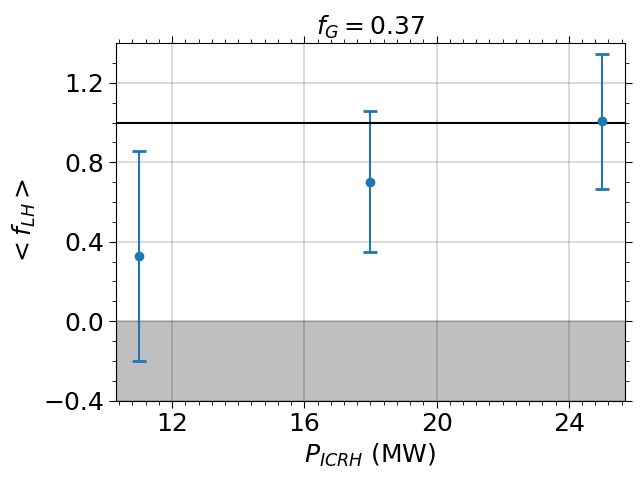}
    \includegraphics[width=0.45\textwidth]{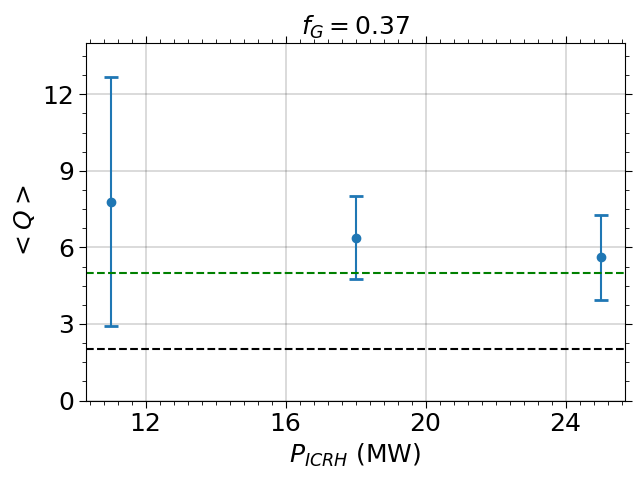}
    \caption{On the left (right) is shown the mean value of $f_{LH}$ ($Q$), together with the standard deviations, which are identified by the error bars, including all the simulations. The gray region indicates unconverged simulations. One can see that increasing the input power the fusion gain is reduced, but $f_{LH}$ increases, showing a better sustained H-mode. The solid black line on the left plot indicates a sustained H-mode condition, according to the Martin scaling. On the right, the green (black) dashed line represents a $Q>5$ ($Q>2$) condition.}
    \label{fig:PRD_mean_std}
\end{figure}

Figure \ref{fig:fW_trends} shows that with the nominal ICRH input power, according to the experimental scalings, the plasma sustain a H-mode only for lower values of W concentration. However, it is worth to remind here that the Martin scaling has very broad error bars, as can be seen in figure \ref{fig:fW_trends}, and it is trained on existing C wall machines, while its reliability for future machines is more uncertain. The LH transition is a complicated phenomenon which involves complex interplay between turbulence suppression, interaction with zonal/sheared flows, global/multi-scale dynamics, formation and steepening of radial electric field gradients and MHD activity \cite{yan_pedestal_2011}, \cite{smith_characterization_2013}, \cite{manz_velocimetry_2014}, \cite{diallo_observation_2014}. Clearly, the possibility to evaluate self-consistently the sustainment of H-mode with high-fidelity simulations which take into account all these ingredients is nearly impractical for the broad database of simulations derived here. This has motivated a simpler approach, based on a scan of the ICRH power for the existing database, with the same randomly assigned input parameters, to assess whether a sustained H-mode, together with high Q values, can be achieved at elevated W concentrations in accordance with existing scalings. Employing PDFs, the results of the power scan are summarized in figure \ref{fig:PRD_PDFs}. Here, the yellow region indicates unconverged simulations.
Figure \ref{fig:PRD_PDFs} shows that the mean value of the fusion power (upper right) is independent of the input power. This is due to the stiffness of the kinetic profiles. The impact of neutron load associated with high fusion power on the superconducting coils of SPARC has not been modeled here, because it is beyond the scope of the paper, but the SPARC design will likely limit fusion to below 140 MW for coil safety limits \cite{creely_overview_2020}. As the ICRH power increases at constant fusion power the fusion gain diminishes, as shown in the upper left plot. Importantly, despite this reduction, all the highest power simulations show $Q>3$, even in the most pessimistic combination of input assumptions. At the bottom of the figure, the PDFs of $f_{LH}$ are shown, calculated using both Martin and Schmidtmayr scaling. No significant difference is observed between the two scalings. However, simulations with the highest ICRH simulations exhibit more robust H-mode sustainment, as the curve is centered around $P_{sep}=P_{LH}$. Therefore, the PRD scenario is capable of sustaining an H-mode configuration with $Q>3$ even at high $f_W$. \\
To summarize the results of the input power scan, the mean value and standard deviation of $Q$ and $f_{LH}$ calculated with Martin scaling, key metrics for evaluating the sustainability and fusion performance, have been calculated. The values are shown in figure \ref{fig:PRD_mean_std}.

\subsection{Pedestal Density Scan}
\begin{figure}[h]
    \centering
    \includegraphics[width=0.45\textwidth]{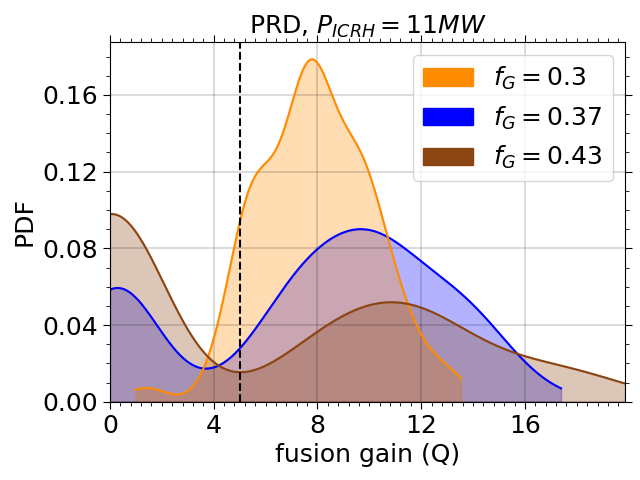}
    \includegraphics[width=0.45\textwidth]{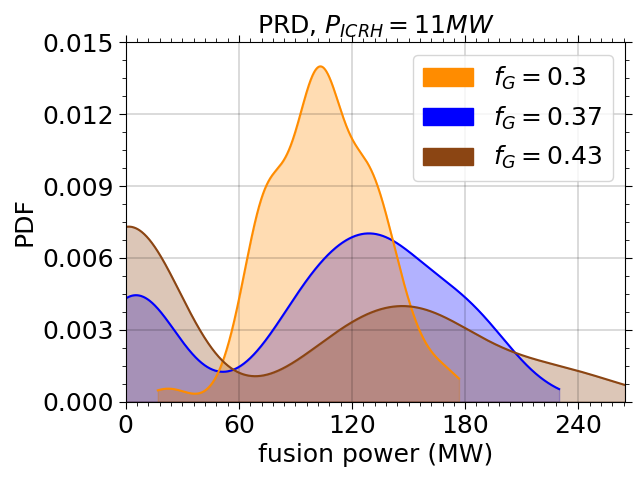}
    \includegraphics[width=0.45\textwidth]{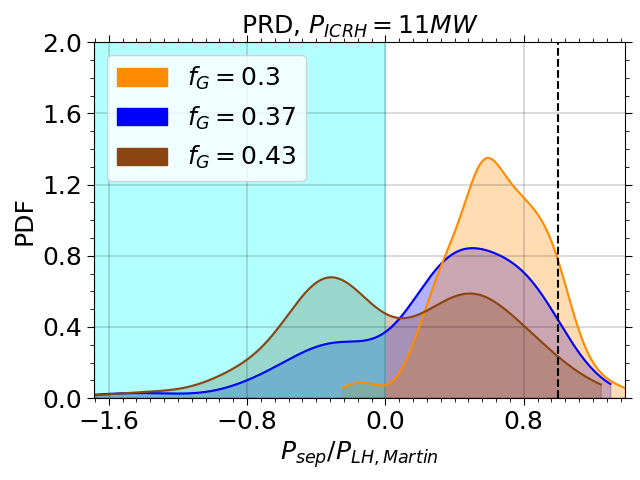}
    \includegraphics[width=0.45\textwidth]{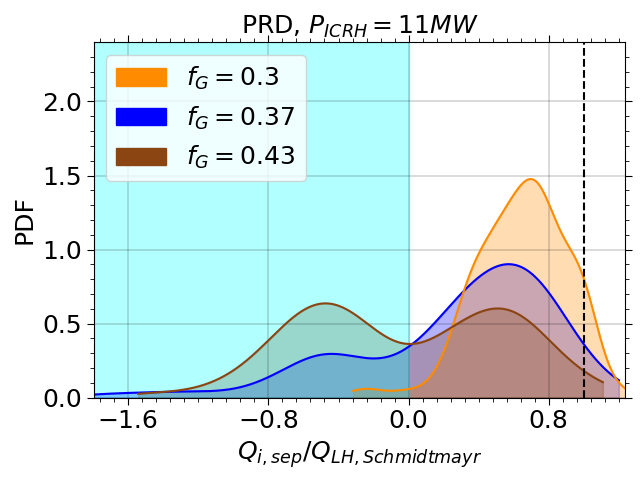}
    \caption{Probability distribution function of different parameters for a scan in density at top of pedestal. Starting from the upper left, clockwise: fusion gain, fusion power, $f_{LH}$ using an ion heat flux threshold, $f_{LH}$ using a total flux threshold. The cyan regions of the plots indicate unconverged simulations. Scaling up the density, the fusion power increases together with the fusion gain, as the input power is fixed. $P_{sep}>P_{LH}$ is rarely met in the database, particularly as density increases. Overall, $Q>4$ is found for all the converged simulations. Starting from the upper left, clockwise the dashed black vertical line indicates: burning plasma condition and sustained H-mode condition with total and ion power scaling.}
    \label{fig:PRD_PDFs_nscan}
\end{figure}

The input power scan described in the previous section has been performed to reach sustained H-mode condition according to the existing scalings. However, plasma density represents another critical parameter during experimental operation. The difficulties related to the modeling of the neutrals penetration and the detailed interaction between them and the background plasma are not treated in this paper, where we simply assume the edge density as a boundary condition, from the top of pedestal to the separatrix. The density affects the core turbulence regime, via $\eta$ (i.e. $\frac{n}{T}\frac{\nabla T}{\nabla n}$) and collisionality (i.e. $\nu_e$), changing the balance between ITG and TEM driven instabilities. Particularly, higher density gradients are a stabilizing mechanism for ITG and a driving term for TEM, while higher collisionality reduces the amount of trapped particle giving overall lower electron heat transport associated with TEM. The equilibration between these two turbulent regimes can impact $\frac{T_i}{T_e}$ in the core and viceversa, affecting the fusion power.\\
Moreover, density influences the overall power balance. In fact, higher densities give overall higher radiation, but, for certain ranges of temperature, also higher fusion power, especially if the density peaking is not affected and the pressure at top of pedestal increases. The value of the pedestal pressure and its trend with the density depend on whether the plasma edge is peeling or ballooning constrained for ELM-y H-mode plasma conditions \cite{connor_magnetohydrodynamic_1998}, \cite{snyder_edge_2002}. Density also impacts H-mode sustainment, as according to the Martin scaling $P_{LH}$ increases with it, indicating that higher edge densities require greater power at the separatrix to sustain H-mode. The determination of H-mode sustainment and performance while scaling the density is not obvious, as it depends on density peaking, turbulent regime, pedestal physics and H-mode sustainment criteria.
This has motivated a density scan, starting from the reference database (i.e. $P_{ICRH}=11MW$ and $f_G=0.37$). In particular, the entire database has been repeated, scaling up and down the density at the pedestal top by 20\%. The results are shown in figure \ref{fig:PRD_PDFs_nscan}.
Within this scan, the density pedestal is scaled up, leading to higher fusion power, as depicted in the upper right figure. As the input power is constant, the fusion gain increases with density. The $f_{LH}$ distribution reveals a weakly improved H-mode sustainment at lower densities. Around the reference PRD operational point, an increase in density results in a higher $P_{LH}$ threshold. However, the effect on the power at the separatrix is non-trivial, due to the interplay between increased radiation and fusion power. Both the $\alpha$ power (i.e. $P_{\alpha}=0.2 P_{fus}$) and $P_{rad,W}$ scale approximately with the square of the density. When their ratio remains constant, the absolute difference between them increases with density (as $P_\alpha>P_{rad}$ for low $f_W$), which leads to higher $P_{sep}$.
\begin{figure}[h]
    \centering
    \includegraphics[width=0.45\textwidth]{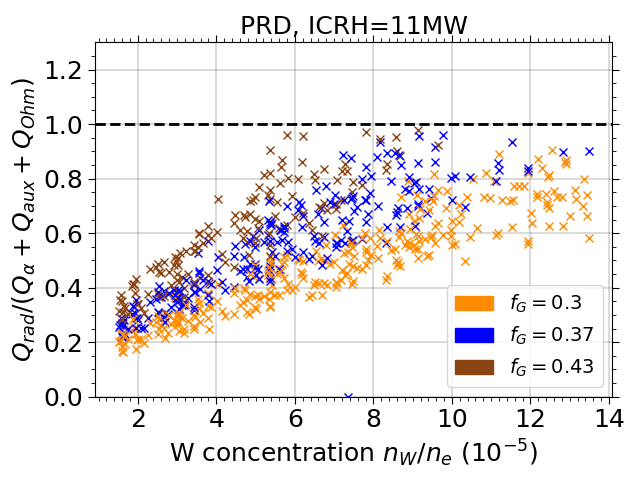}
    \includegraphics[width=0.45\textwidth]{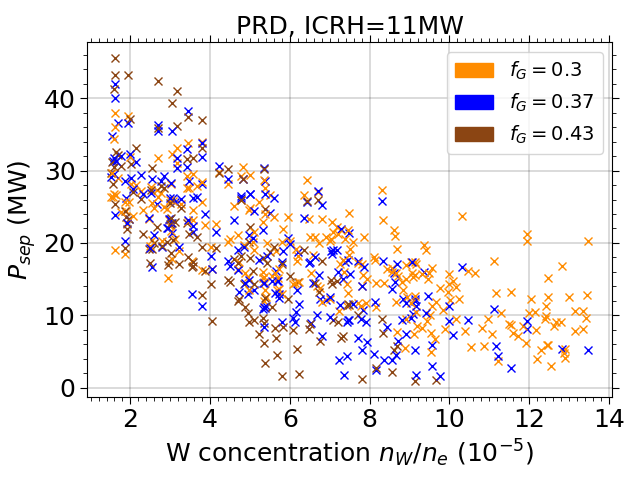}
    \includegraphics[width=0.45\textwidth]{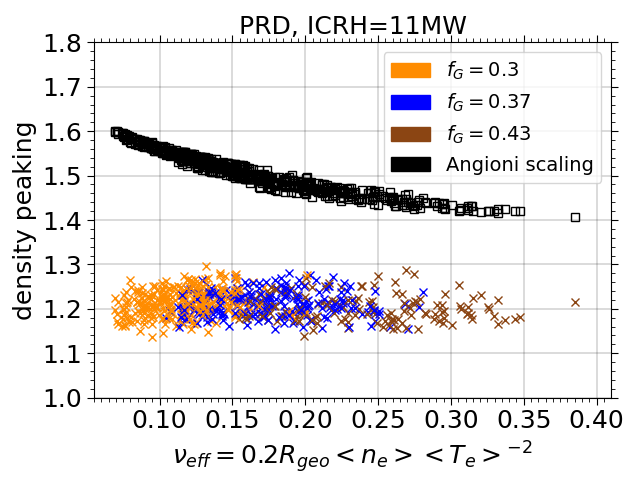}
    \includegraphics[width=0.45\textwidth]{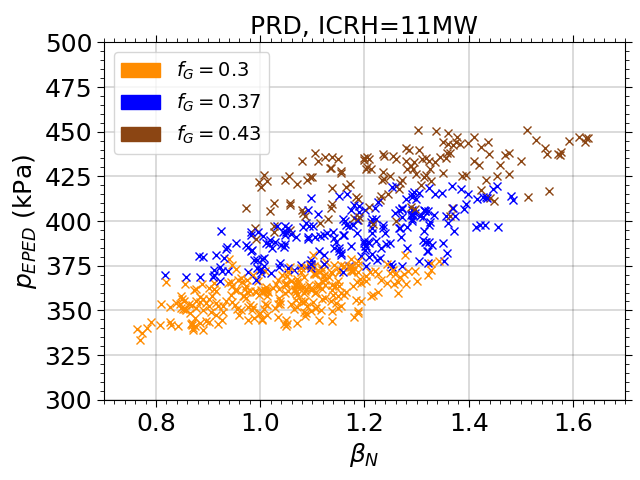}
    \caption{Top left: the ratio between W radiation and fusion+auxiliary+ohmic power vs the W concentration for 3 different values of density at top of pedestal. This ratio is higher for higher densities at constant $f_W$, therefore the increased fusion power is balanced by higher radiation, leading to a constant $P_{sep}$. Top right: power at the separatrix vs W concentration for the density scan. No clear trend is observed with changing density. Bottom left: density peaking, defined as $n_e(\Psi_n=0.2)/\langle n_e\rangle$, where $\Psi_n$ is the normalized poloidal flux, together with the value calculated from the Angioni scaling \cite{angioni_scaling_2007}, with respect to which TGLF shows lower values. A very weak dependence of the peaking is shown with respect to the collisionality and pedestal density, meaning that when the pedestal is scaled up the entire profile scales up around the same order within the core, increasing the fusion power accordingly. Bottom right: the linear relationship between $\beta_N$ and the predicted pressure from EPED-NN is shown for the density scan.}
    \label{fig:PRD_Qrad_nu}
\end{figure}
However, figure \ref{fig:PRD_Qrad_nu} demonstrates that the ratio of $Q_{rad}$ (i.e. volume integral of $P_{rad}$) to the total input power (i.e. $Q_{input}$, the sum of $\alpha$, ohmic and auxiliary power) increases with density. This results in a constant value of $Q_{input}-Q_{rad}$, which keeps $P_{sep}$ nearly unchanged across different density levels (see upper right plot in figure \ref{fig:PRD_Qrad_nu}). It is important to note that the pedestal density scan is limited in range, leading to minimal differences in H-mode sustainment. The higher $P_{fus}$ is primarily due to the increased density and the relatively unchanged peaking of the density profile, as shown in figure \ref{fig:PRD_Qrad_nu}.
\begin{figure}[ht]
    \centering
    \includegraphics[width=0.45\textwidth]{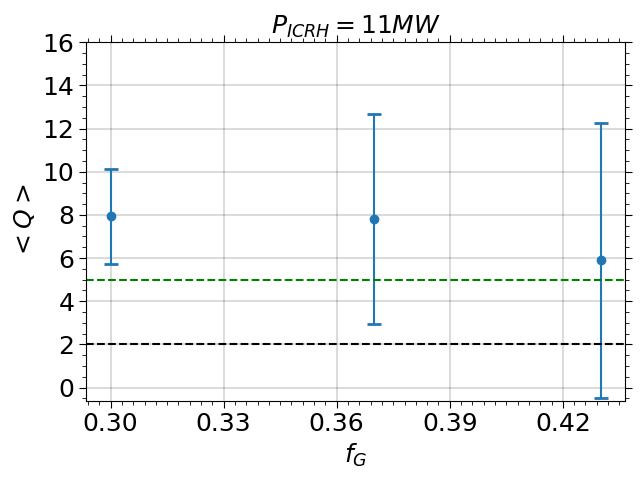}
    \includegraphics[width=0.45\textwidth]{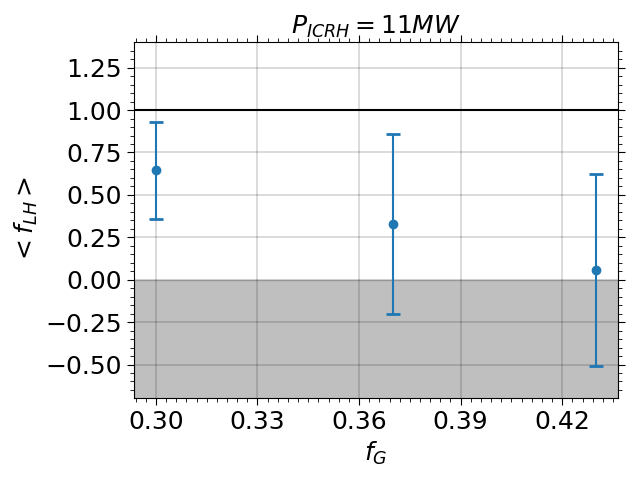}
    \caption{On the left (right) is shown the mean value of $Q$ ($f{LH}$), together with the standard deviations, which are identified by the error bars, including all the simulations. One can see that increasing the density the fusion gain is increased, due to higher fusion, but the confidence in the sustainment of the H-mode regime is reduced ($f_{LH}<1$). The solid black line on the right plot indicates a sustained H-mode condition, according to the Martin scaling, while the gray region indicates unconverged conditions. On the left the green (black) dashed line represents the burning plasma ($Q>2$) condition.}
    \label{fig:PRD_mean_std_nscan}
\end{figure}
This indicates that the entire profile shifts upward, generating more fusion. Additionally, the pedestal pressure calculated by EPED-NN scales linearly with $\beta_N$, which is proportional to the density in this region of the operational space, as shown at the bottom of figure \ref{fig:PRD_Qrad_nu}. The bottom left plot also reveals that the density peaking predicted by TGLF-SAT2 does not significantly change with collisionality. This is in contrast with the Angioni scaling \cite{angioni_scaling_2007}, calculated as $\nu_{Angioni}=1.347-0.117\log(\nu_{eff})-4.03\beta$, where the definitions of $\nu_{eff}$ and $\beta$ can be found in \cite{angioni_scaling_2007}. The Angioni scaling, plotted in black for reference, predicts higher density peaking values than those observed in the database, which align with prior high-fidelity nonlinear CGYRO simulations of the PRD \cite{rodriguez-fernandez_nonlinear_2022}. However, the runs performed in this paper showed a peaking consistent with what was found by TGLF simulations in \cite{rodriguez-fernandez_nonlinear_2022}, and a lower peaking is conservative in terms of fusion performance. Moreover, the database used to derive the Angioni scaling exhibits substantial scatter with respect to the experimental data, particularly within the range of density peaking $\nu_{n_e}=[1.2-1.6]$. This suggests that TGLF-predicted outliers may still fall within experimentally observed values.
To summarize the results of the density scan, figure \ref{fig:PRD_mean_std_nscan} presents the average value of Q and $f_{LH}$ across different volume-averaged Greenwald fractions (i.e. $f_G$), along with their respective standard deviations, represented by the error bars. Lower densities reduce the uncertainty on Q and improve the sustainment of H-mode, due to the smaller impact of radiation on performance.

\subsection{TGLF spectrum analysis}
As previously mentioned, changing ion to electron temperature ratio, density, collisionality and other plasma parameters can affect the turbulence.
\begin{figure}[h]
    \centering
    \includegraphics[width=0.4\textwidth]{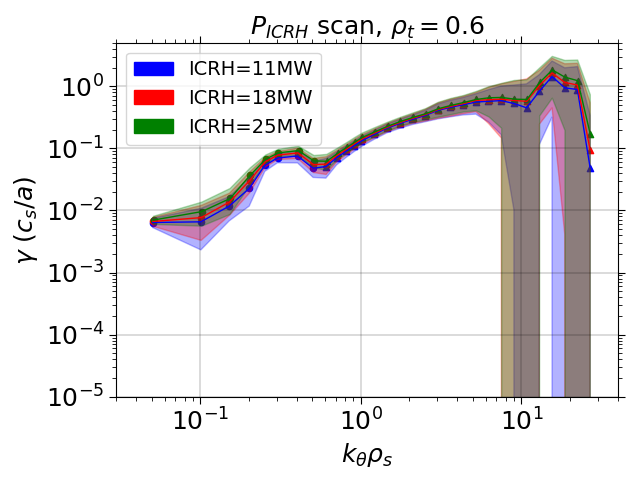}
    \includegraphics[width=0.4\textwidth]{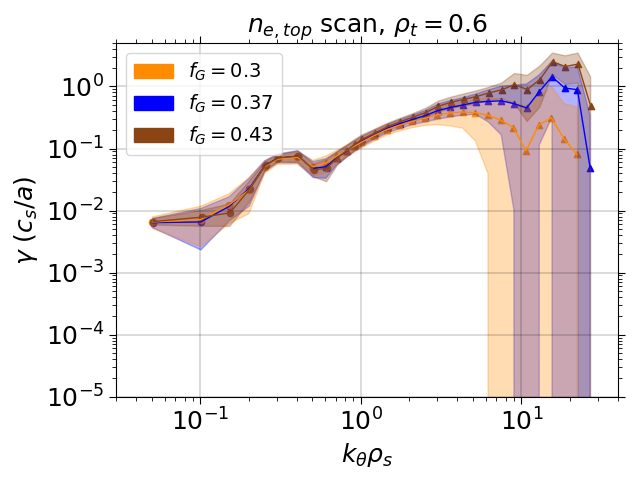}
    \includegraphics[width=0.4\textwidth]{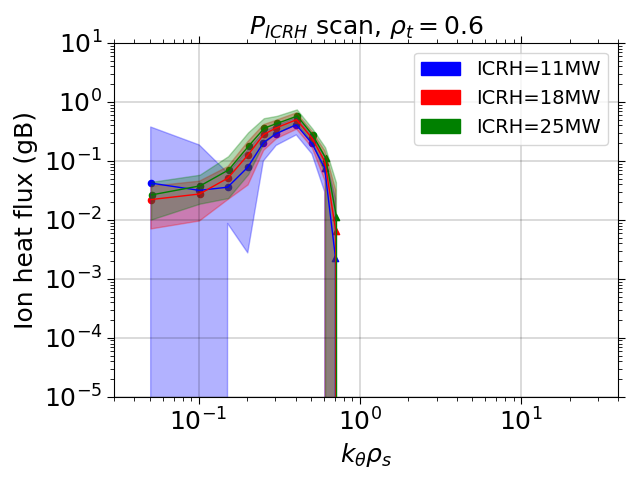}
    \includegraphics[width=0.4\textwidth]{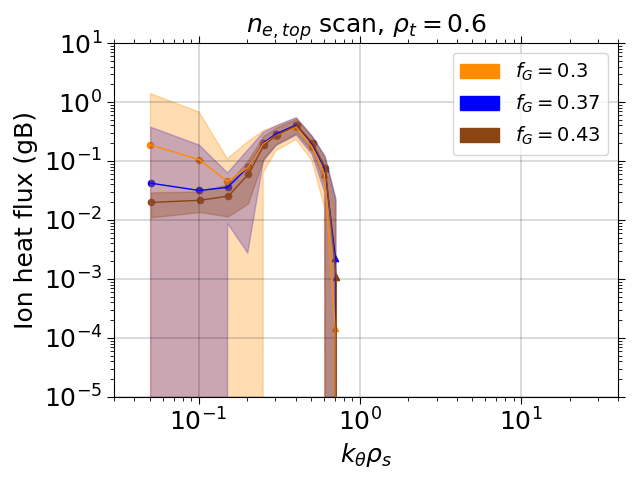}
    \includegraphics[width=0.4\textwidth]{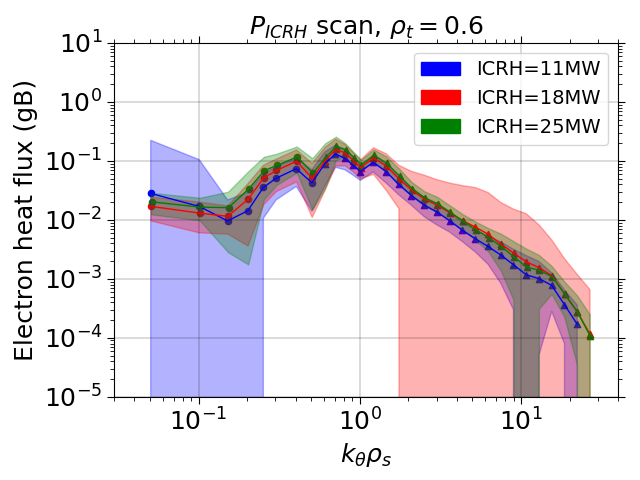}
    \includegraphics[width=0.4\textwidth]{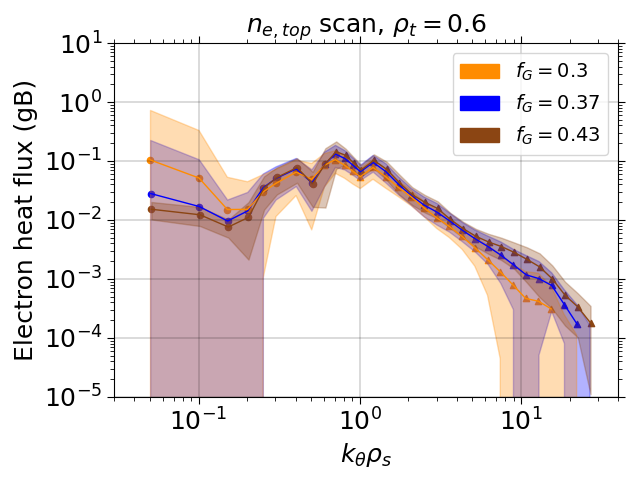}
    \caption{From top to bottom on the left (right) the average TGLF growth rates, ion and electron heat flux spectrum at $\rho_t=0.6$ for the power (density) scan, together with their standard deviations, quantified by the error bars. The growth rates increase slightly with the ICRH power, due to the small role played by $P_{ICRH}$ compared to $P_{fus}$ in the power balance, while higher densities drive more unstable modes at high-k electron scales, associated with an increased electron heat flux. The dots represent ion-directed modes, while the triangles electron-directed modes. The low-k part of the spectrum shows robust ITG dominance.}
    \label{fig:PRD_growth_rates}
\end{figure}
Therefore, the turbulent spectrum of TGLF should be analyzed. Considering the large amount of simulations, the analysis should be performed via statistical methods, to extract trends and global information.
The mean values of growth rates, electron and ion heat fluxes at $\rho_t=0.6$, where $\rho_t$ is the square root of the normalized toroidal magnetic flux, are shown in figure \ref{fig:PRD_growth_rates} for both the power and density scans, together with their standard deviation, represented by error bars. In the following "low-k" will refer to the region of the spectra where $k_y\rho_s<1$, while "high-k" to the $k_y\rho_s>1$ region. "Middle-k" refer to electron-scale regions between $k_y\rho_s=1$ and $k_y\rho_s=2$. Here $k_y$ refers to the poloidal wave number of the fluctuation and $\rho_s$ is the ion sound speed Larmor radius. In figure \ref{fig:PRD_growth_rates} the dots represent ion-directed instabilities, while the triangles electron-directed ones. The figure shows that at low-k the ion-directed modes are prevalent, meaning the ITG dominates over TEM at these scales. This dominance has been confirmed observing that the ion heat flux scales mainly with $a/L_{T_i}$, where $L_{T_i}$ is the ion temperature gradient length, and is primarly reduced by $\beta_e$.
\begin{figure}[h]
    \centering
    \includegraphics[width=0.45\textwidth]{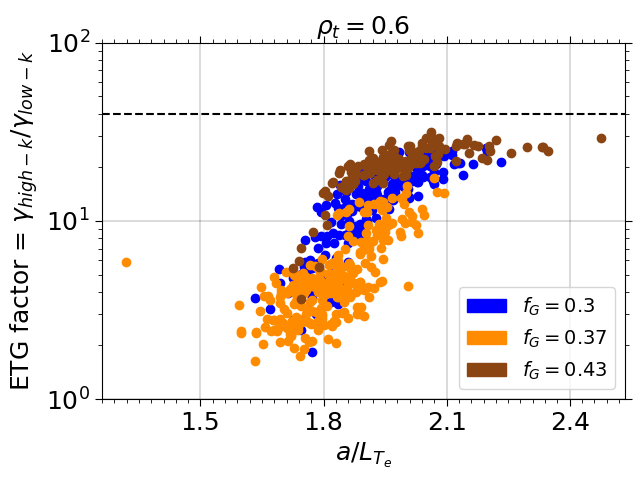}
    \includegraphics[width=0.45\textwidth]{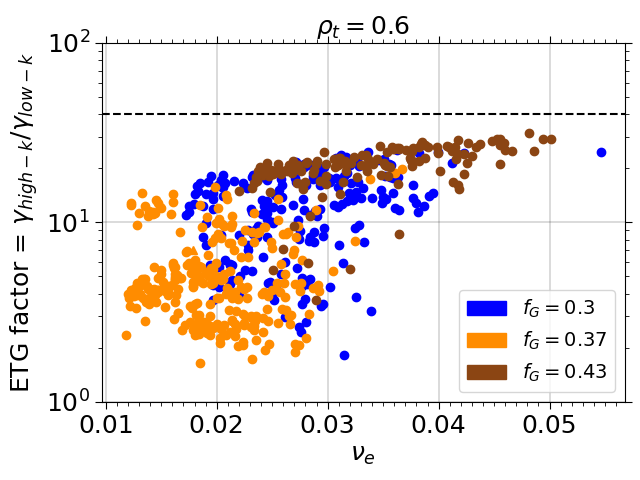}
    \includegraphics[width=0.45\textwidth]{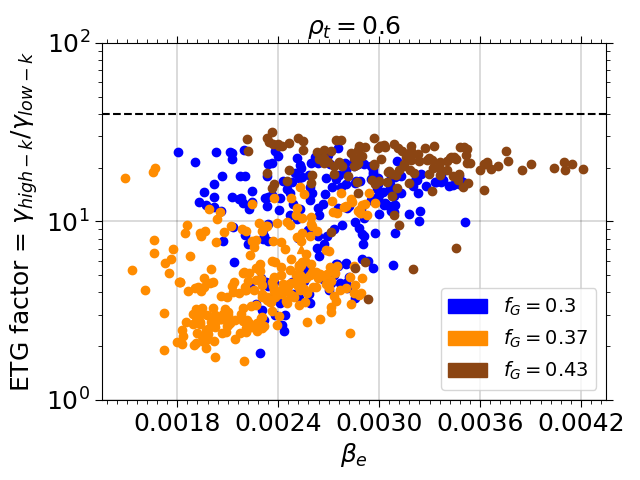}
    \includegraphics[width=0.45\textwidth]{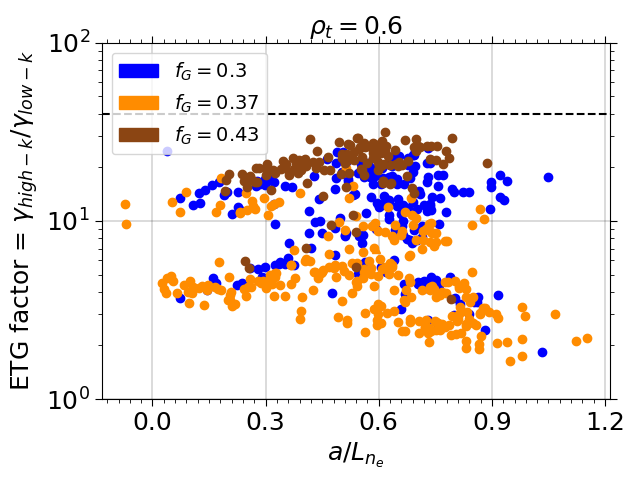}
    \caption{From top left, clockwise: ETG factor (i.e. $f_{ETG}=\gamma_{high-k}/\gamma_{low-k}$) as function of $a/L_{T_e}$, $\nu_e$, $a/L_{n_e}$ and $\beta_e$ at $\rho_t=0.6$, for the pedestal density scan. The strong correlation between $f_{ETG}$ and most of these parameters indicates an increased ETG and reduced ITG/TEM activity for higher densities. However, all the simulations lay below $f_{ETG}=40$ (dashed black line), suggesting that the ETG activity is always marginal.}
    \label{fig:PRD_ETG_factor}
\end{figure}
The power scan does not show significant modifications of the spectra, because the change in $P_{ICRH}$, consistently with stiff transport, does not affect the profiles and the heating is predominantly driven by $\alpha$ particles for the burning plasma conditions of the analyzed scenario. Increasing the density drives more high-k unstable modes, which results in slightly higher electron heat fluxes at high electron scales. This behaviour is associated with an increase in ETG or middle-k TEM activity for higher density cases.
In order to address this, an ETG factor (i.e. $f_{ETG}$), defined in line with \cite{howard_multi-scale_2016-1} as $\gamma_{high-k}/\gamma_{low-k}$, where $\gamma_{high-k}$ and $\gamma_{low-k}$ are respectively the highest growth rates for high- and low-k, has been analyzed and plotted with respect to TGLF input variables. The plots are shown in figure \ref{fig:PRD_ETG_factor}.
While no relevant trend of $f_{ETG}$ is found with respect to $Z_{eff}$, $T_i/T_e$ and $a/L_{T_i}$, due to the small variations of these parameters across the various simulations, strong relations are found with respect to $\beta_e$, $\nu_e$, $a/L_{T_e}$, and a weak relation is found with respect to $a/L_{n_e}$. In particular $f_{ETG}$ increases with $a/L_{T_e}$, $\beta_e$ and $\nu_e$. This behaviour is associated with increased ETG and reduced TEM/ITG activity, because the ETG main drive is $\nabla T_e/T_e$, high $\beta_e$ stabilizes ITGs and high $\nu_e$ stabilizes TEMs. The effect of the stabilizing terms can also be seen in the middle and bottom right plots in figure \ref{fig:PRD_growth_rates}, where the ion and electron heat fluxes at low-k are reduced at higher densities. The dependence of $f_{ETG}$ on $a/L_{n_e}$ is weak because the normalized density gradient represents a driving term for TEMs, but a stabilizing term for ITGs, averaging out its effect on the heat flux at low-k scales. Despite an increased ETG activity is predicted by TGLF, its overall contribution to the total flux is marginal. In fact, $f_{ETG}<40$ for all the converged simulations. This threshold has been found for Alcator C-mod multi-scale gyrokinetic simulations in \cite{howard_multi-scale_2016-1}, as rule of thumb to determine ETG effect when combined with low-k turbulence. This reference value is represented by the dashed lines in figure \ref{fig:PRD_ETG_factor}.

\subsection{Effect of high-k turbulence}
\begin{figure}[h]
    \centering
    \includegraphics[width=0.45\textwidth]{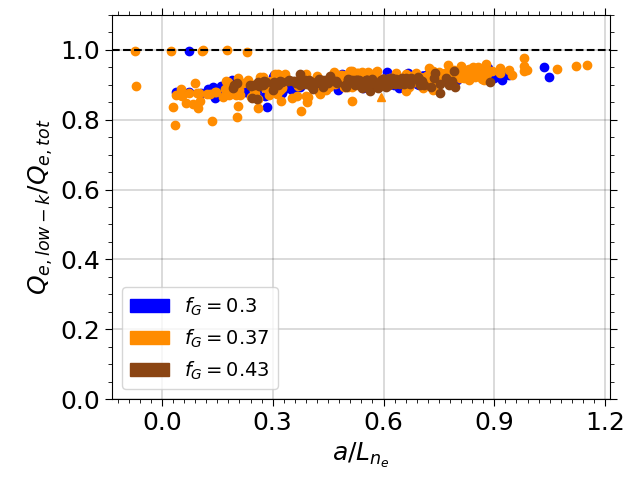}
    \includegraphics[width=0.45\textwidth]{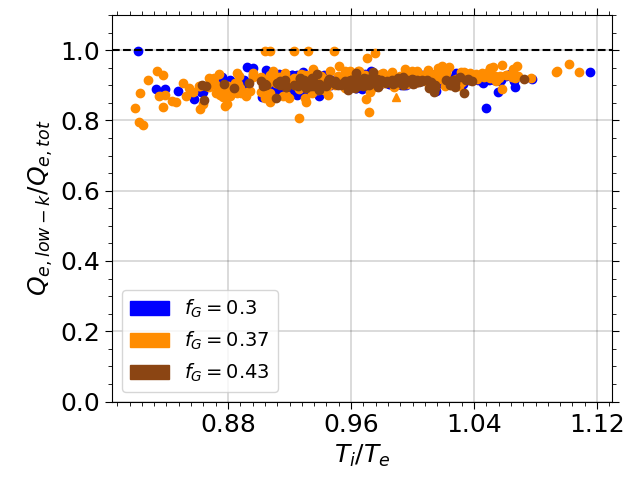}
    \caption{On the left (right) the ratio of the low-k to total electron heat flux vs the normalized density gradient ($T_{i}/T_{e}$). The weak linear relationship with both the variables suggests that $Q_{e,low-k}$ is impacted by TEM activity. The dashed black line indicates pure low-k electron heat flux. $Q_{e,low-k}$ gives always most of the contribution to the total $Q_e$.}
    \label{fig:qe_ratio_nscan}
\end{figure}

In order to investigate further the role played by high-k activity, we analyzed the ratio between the low-k and total electron heat transport. This ratio is shown in figure \ref{fig:qe_ratio_nscan}.
\begin{figure}[htbp]
    \centering
    \includegraphics[width=0.4\textwidth]{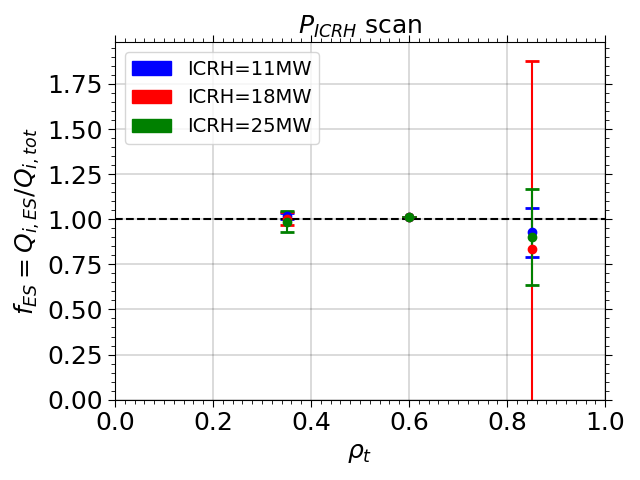}
    \includegraphics[width=0.4\textwidth]{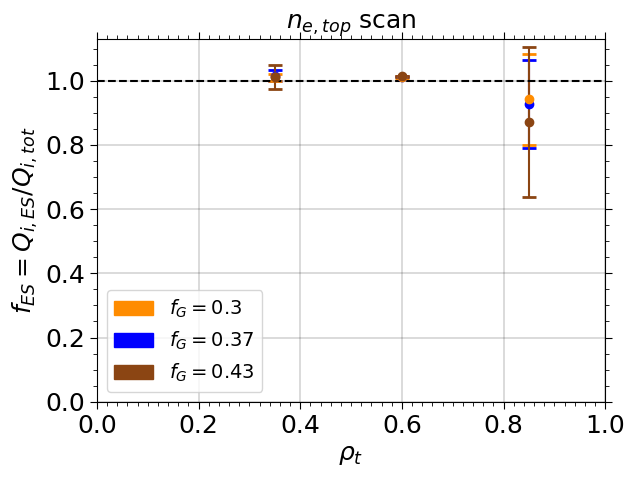}
    \includegraphics[width=0.4\textwidth]{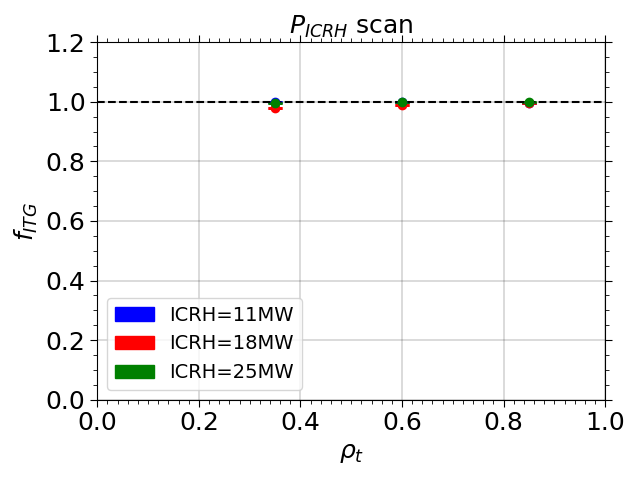}
    \includegraphics[width=0.4\textwidth]{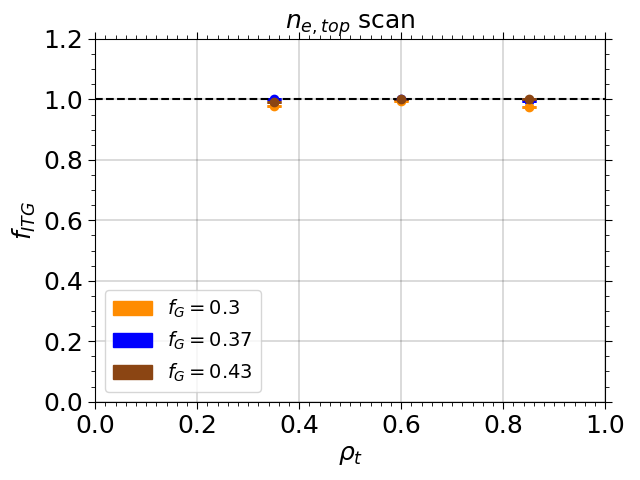}
    \includegraphics[width=0.4\textwidth]{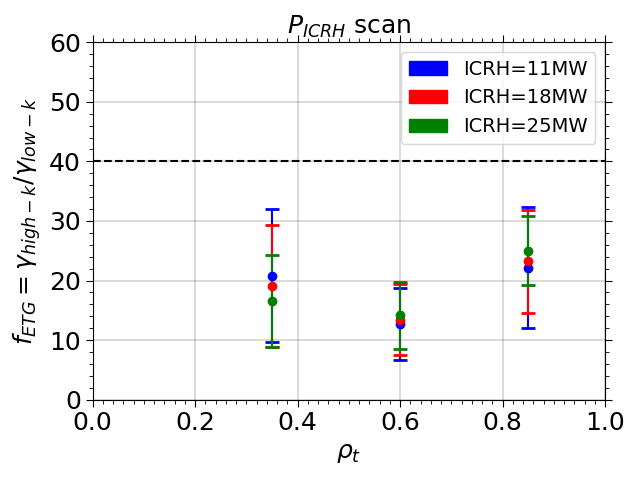}
    \includegraphics[width=0.4\textwidth]{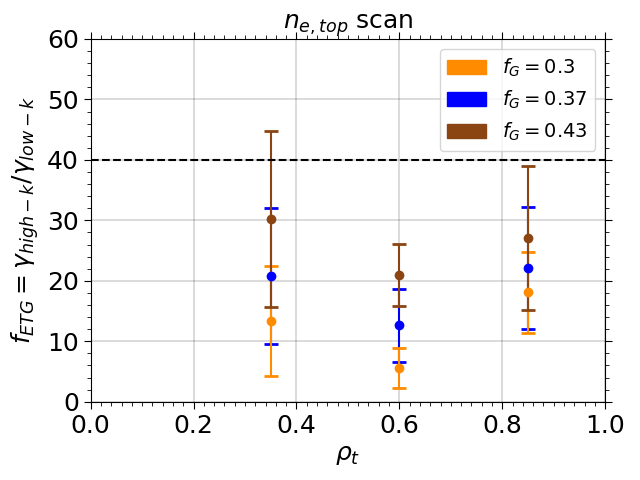}
    \includegraphics[width=0.4\textwidth]{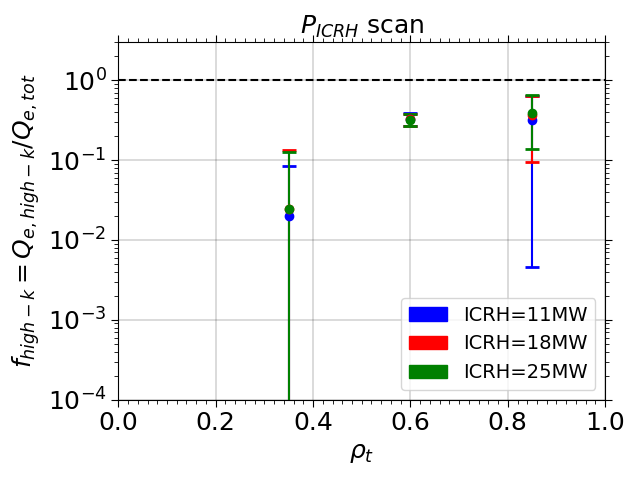}
    \includegraphics[width=0.4\textwidth]{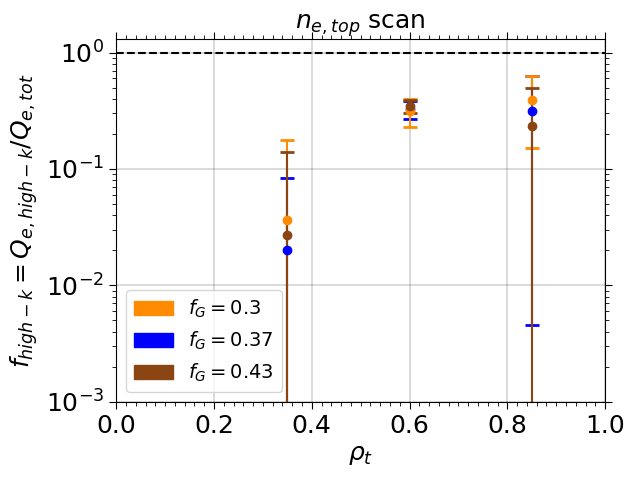}
    \caption{From top to bottom on the left (right) the mean values and standard deviations of $f_{ES}$, $f_{ITG}$, $f_{ETG}$ and $f_{high-k}$ for 3 different radial locations for the ICRH power scan (pedestal density scan). The ion heat flux is robustly found electrostatic, the ion-scale is completely ITG dominated and a low ETG activity is found overall, with a slight increase for higher densities. The contribution of high electron scales to the total electron flux rarely reaches 50\%.}
    \label{fig:PRD_spectra_global}
\end{figure}
This ratio increases with the normalized density gradient and $T_{i}/T_{e}$, which are conditions in which the TEM regime increases and ITG is suppressed. This suggests that the electron heat flux at high-k is associated with middle-k TEM activity. In fact, redefining low-k as the range where $k_y\rho_s<2$ and high-k as the range where $k_y\rho_s>2$ \cite{waltz_coupled_2007}, the average $Q_{e,low-k}/Q_{e,tot}$ goes from 0.65 to 0.92. This suggests that most of the flux associated with earlier defined high-k region arises from mid-k TEM activity. \\
In order to assess the effect of electro-magnetic instabilities, the total ion heat flux and its electro-static part have been compared. The flux is found almost purely electro-static with a maximum of 5\% deviation at high $\beta_e$. \\
The TGLF turbulence spectra exhibited insensitivity to the ICRH input power; therefore, the detailed analysis primarily focused on the spectra obtained from the pedestal density scan at $\rho_t=0.6$. However, to ensure that the observed plasma behavior is consistent across different radial locations, additional analyses were conducted at $\rho_t=0.35$ and $\rho_t=0.85$. The former lies within the typical inversion radius for SPARC PRD plasmas, while the latter is closer to the pedestal top position, characterized by lower gradients and higher collisionality. Given the large size of the database, a statistical approach was employed to summarize the data. Specifically, the mean values and standard deviations of four key variables were plotted for both the ICRH power and density scans in figure \ref{fig:PRD_spectra_global}. These variables are:
\begin{itemize}
    \item $f_{ETG}$, the factor indicating the ETG activity,
    \item $f_{ES}$, the electrostatic fraction of ion heat flux (i.e. $Q_{i,ES}/Q_{i,tot}$),
    \item $f_{ITG}$, the percentage of simulations where the ion heat flux is dominated by ion-directed instabilities,
    \item $f_{high-k}$, the fraction of high-k electron heat flux (i.e. $Q_{e,high-k}/Q_{e,tot}$)
\end{itemize}
The results reveal that the ion heat flux remains almost entirely electrostatic for both scans, with the electromagnetic (EM) contribution increasing slightly at $\rho_t=0.85$. However, even at this location, the mean EM flux contribution does not exceed 25\%. 
The $f_{ITG}$ metric consistently equals 1 across all ICRH powers, pedestal densities, and radial locations, indicating that ion temperature gradient (ITG) instabilities are the dominant mode across the explored parameter space.  While no clear trend in $f_{ETG}$ is found with radial location,an increase in pedestal density consistently leads to a rise in this factor. This indicates enhanced linear high-k electron-scale activity at higher densities. However, the mean $f_{ETG}$ is consistently lower than 40, and this increased linear instability is not associated with a significant rise in high-k electron heat flux, as previously demonstrated for $\rho_t=0.6$. Consequently, $f_{high-k}$ remains nearly constant for both ICRH and density scans at fixed radial location. However, $f_{high-k}$ at $\rho_t=0.35$ is an order of magnitude lower than at the other locations. This difference is attributed to increased low-k trapped electron mode (TEM) activity in the inner core, where reduced collisionality prevails.

\section{SPARC 8T H-mode scenario}
\begin{figure}[h]
    \centering
    \includegraphics[width=0.45\textwidth]{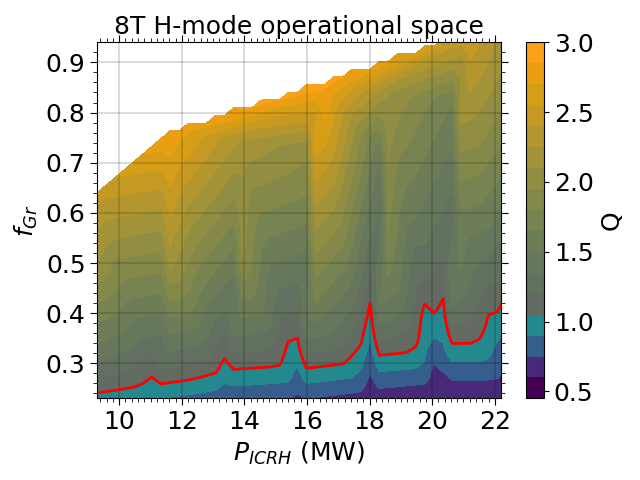}
    \includegraphics[width=0.45\textwidth]{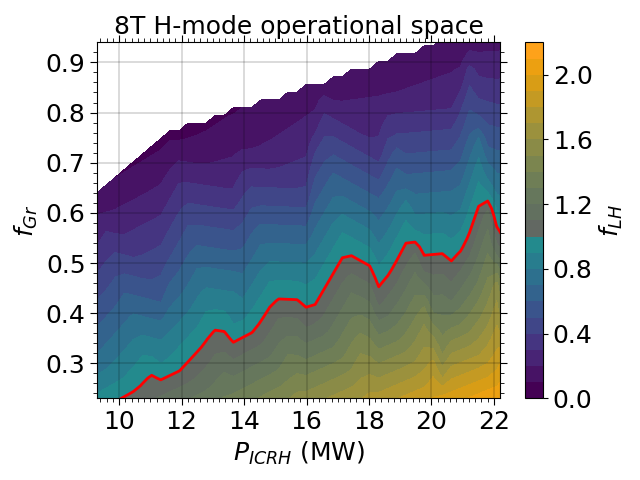}
    \includegraphics[width=0.45\textwidth]{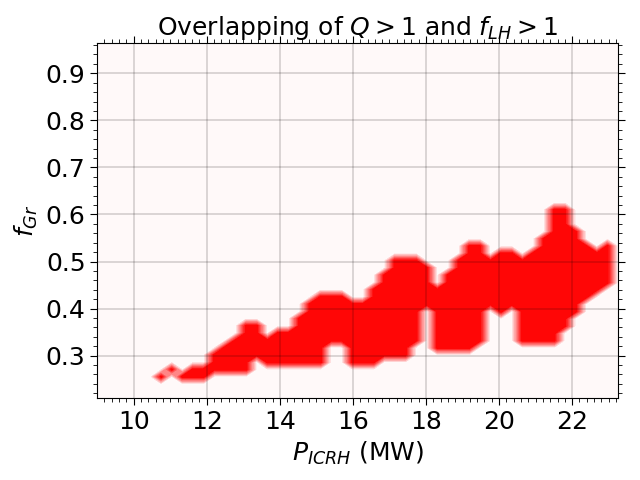}
    \caption{At the top left (right) $Q$ ($f_{LH}$) as function of $f_G$ and $P_{ICRH}$ for the SPARC 8T H-mode. The red lines show transition to $Q>1$ and $f_{LH}>1$. At the bottom, in red is the overlapping region of $Q>1$ and $f_{LH}>1$, where Q values between 1 and 1.6 are found. Here, an operational point which is breakeven-relevant and H-mode sustained has been found for maximum $P_{ICRH}$ and $f_G=0.46$, showing Q=1.4.}
    \label{fig:8T_scoping}
\end{figure}

\begin{figure}[t]
    \centering
    \includegraphics[width=0.4\textwidth]{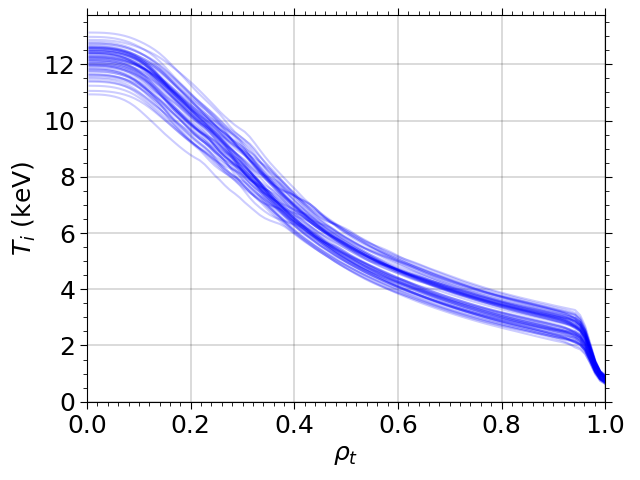}
    \includegraphics[width=0.4\textwidth]{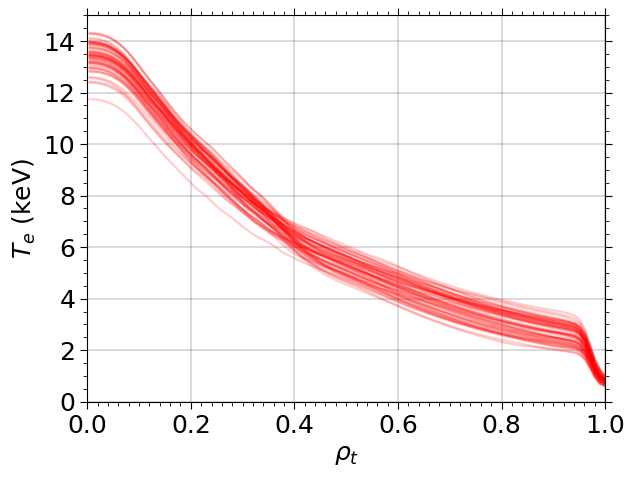}
    \includegraphics[width=0.4\textwidth]{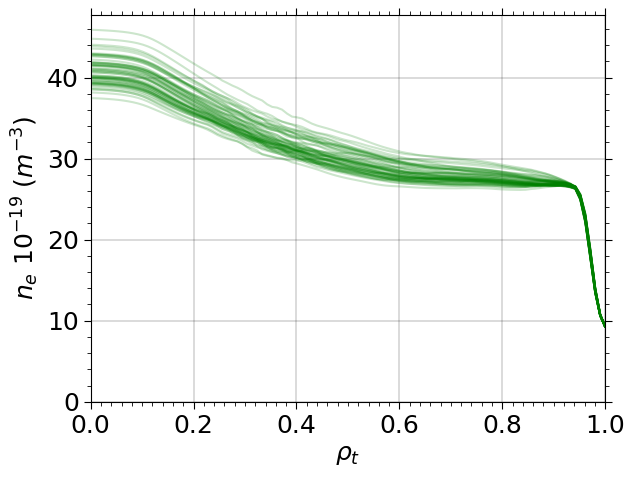}
    \caption{SPARC 8T-Hmode converged kinetic profiles with $f_G=0.46$ and $P_{ICRH}=25MW$.}
    \label{fig:8T_kinetic_profiles}
\end{figure}

While the PRD has been widely analyzed and the prediction of its performance has been quantified with low and high-fidelity models \cite{creely_overview_2020}, \cite{rodriguez-fernandez_overview_2022}, the reduced field (8T) scenario is still quite unexplored. The 8T H-mode employs a different heating scheme with respect to PRD for ICRH power, using H minority species. This scenario is very attractive for many reasons, and deserves a broader scoping of its features and performance. In fact, at reduced magnetic field, the LH power threshold is reduced, according to the existing scalings. A reduced magnetic field implies lower plasma current to maintain a fixed safety factor, avoiding critical MHD activity and/or disruption. With lower $B_t$ and $I_p$, the combined effect on $E\times B$ shear and gyroBohm transport scaling amplifies the transport \cite{angioni_dependence_2023}. While this characteristic may result in an overall lower fusion performance, it also reduces the plasma energy content. This, in turn, lowers the technological risk associated with potential disruptions, as the resulting damage may be more easily mitigated. This feature makes the 8T scenario particularly attractive for the early phase of SPARC operation, where minimizing damage to plasma-facing components (PFCs) is likely to take precedence over achieving very high fusion performance. The main parameters of the 8T H-mode scenario are listed in table \ref{tab:8T}.
\begin{table}[h]
  \centering
  \caption{List of the essential engineering parameters of the SPARC 8T H-mode discharge.}
  \begin{tabular}{cc}
    \toprule
    \rowcolor{lightblue} % Row color for header
    \textbf{Parameter} & \textbf{Value} \\
    \midrule
          $B_t$  &  8 T  \\
          $I_p$  &  5.7 MA  \\
          $k_{sep}$  &  2  \\
          $\delta_{sep}$  &  0.54  \\
          $q^*_{Uckan}$  &  3  \\
    \bottomrule
  \end{tabular}
  \label{tab:8T}
\end{table}
To run the reference 8T H-mode database, the same framework described earlier was used, with the only difference being the heating deposition profiles, which were derived from a TRANSP+TORIC simulation with H minority heating. The He3 species was then replaced with the same concentration of H (0.05) in ASTRA, while maintaining the same assumptions about W and lumped impurity species and their concentrations. This adjustment led to an increase in the $f_{DT}$ standard assumption to ensure quasi-neutrality. The reference concentrations for all species are listed in Table \ref{tab:8T_concentrations}.
\begin{table}[h]
  \centering
  \caption{Standard assumptions of species concentrations for the SPARC 8T H-mode.}
  \begin{tabular}{cccc}
    \toprule
    \rowcolor{lightblue} % Row color for header
    \textbf{$f_{DT}$} & \textbf{$f_H$} & \textbf{$f_W$} & \textbf{$f_{lumped}$} \\
    \midrule
            $0.9$ & $0.05$ & $1.5\cdot 10^{-5}$ & $6.2\cdot 10^{-3}$        \\
    \bottomrule
  \end{tabular}
  \label{tab:8T_concentrations}
\end{table}
Considering the discussed aspects, it is important to explore the operational space to find
an optimal point that maximizes the fusion gain and assure a sustained H-mode with a certain degree of fidelity. This has been done by spanning the $f_G$ (i.e. density, at fixed current) vs $P_{ICRH}$ space. These parameters can usually be controlled during the operation of a tokamak, and they strongly affect the performance, as seen in the previous section. The results of this scoping are shown in figure \ref{fig:8T_scoping}.
\begin{figure}[t]
    \centering
    \includegraphics[width=0.45\textwidth]{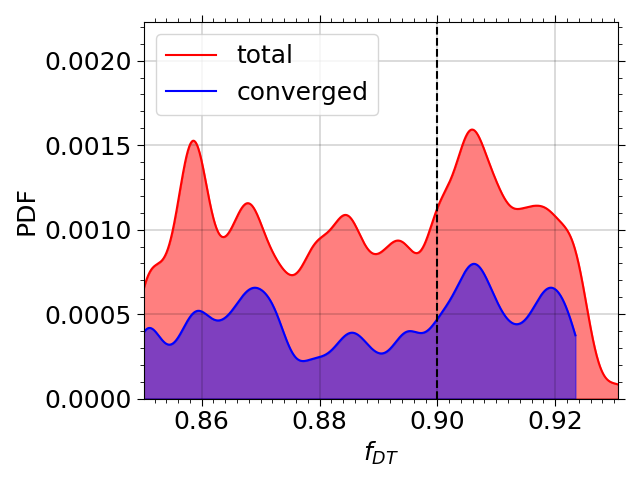}
    \includegraphics[width=0.45\textwidth]{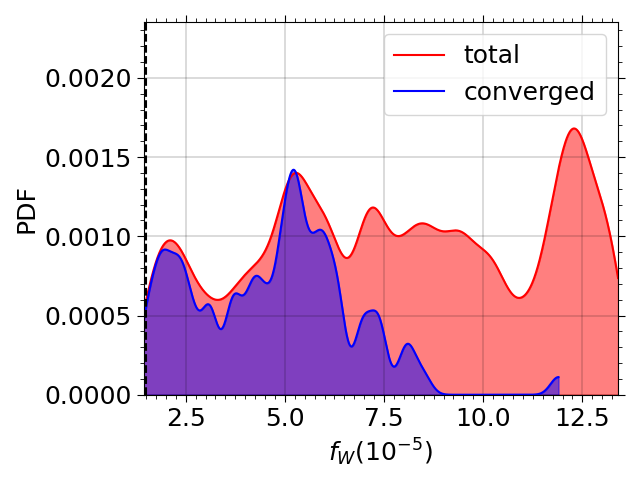}
    \includegraphics[width=0.45\textwidth]{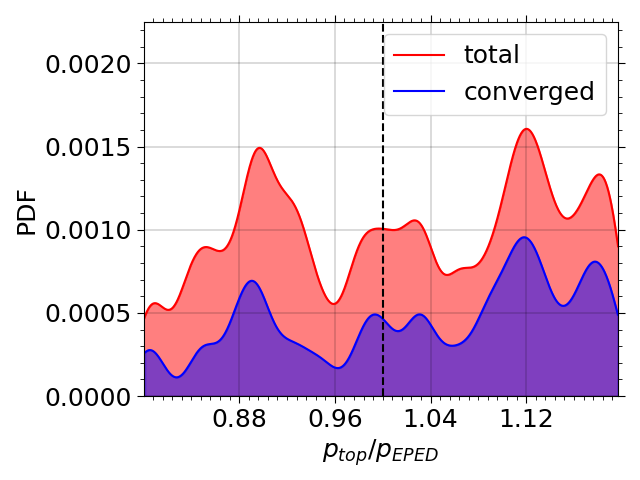}
    \includegraphics[width=0.45\textwidth]{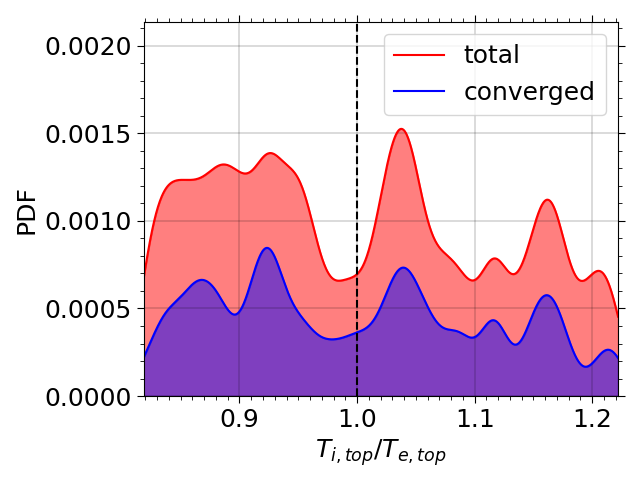}
    \caption{PDFs of the random input assumptions for the 8T H-mode. An asymmetry is found for $f_W$, due to the impact of W radiation.}
    \label{fig:8T_input_PDFs}
\end{figure}
Here $Q$ and $f_{LH}$ are shown as main metrics to measure the performance of the discharge. While the fusion power has not been prioritized, $Q>1$ has been selected as the criterion to evaluate the quality of fusion performance, in line with a strategy that considers this scenario as relevant for SPARC's early operation. The $Q>1$ condition, which is often met, is above the red line in the brighter region of top left plot. A harder condition to obtain is $f_{LH}>1$ (i.e. sustained H-mode), where the LH power threshold is computed with the Martin scaling \cite{martin_power_2008}. This condition, below the red line in the top right plot, is fulfilled in a narrower region at low densities and high input powers, in agreement with the $f_{LH}$ trends observed in the previous section for the PRD. Overlapping the regions of $Q>1$ and $f_{LH}>1$, an optimal operational point at maximum $P_{ICRH}$ and $f_G=0.46$ has been identified. However, to take into account more pessimistic W concentration and be conservative about the H-mode sustainment, an operational point with $P_{ICRH}=25MW$ has been selected. Starting from this condition and the previously listed reference input values, a sensitivity study of performance has been conducted, permuting the values of $f_W$, $f_{DT}$, $T_{i,top}/T_{e,top}$ and $p_{top}/p_{EPED}$ within the same ranges as in the PRD (except for the DT concentration, which was shifted up by 5\%). The uniform variations of the input parameters are shown in figure \ref{fig:8T_input_PDFs}, with the same logic as in figure \ref{fig:PRD_ranges}.
The red curve represents the total simulations, while the blue one is related to the converged cases. An asymmetry in $f_W$ distribution is found, due to the W radiation. In the figure, although the full range of input values is covered, the distributions are not flat due to insufficient sampling density. In the future, an extension of the database, including more input values, is planned. With the shown random permutations, an initial database of 192 simulations has been obtained. Of these, 54\% survived the radiation effect, which is a lower percentage compared to the full-field H-mode. This is because, at lower fields, the pedestal and fusion power are lower, making radiation more competitive with the $\alpha$ power, leading then more easily to a radiative collapse or unsustained H-mode. The trends of Q and $f_{LH}$ with respect to W concentration, respectively insensitive and inversely proportional, are very similar to those found in figure \ref{fig:fW_trends}, and thus have not been repeated here. Additionally, as with the PRD, the most impactful input variables on the performance are $T_{i,top}/T_{e,top}$ and $p_{top}/p_{EPED}$, both of which show a direct proportionality with Q. Its trend with these 2 parameters is the same as in figure \ref{fig:Q_trends}, so the authors have opted not to show them to avoid redundancy. The kinetic profiles of the database are shown in figure \ref{fig:8T_kinetic_profiles}.

\subsection{Pedestal Density Scan}
\begin{figure}[h]
    \centering
    \includegraphics[width=0.45\textwidth]{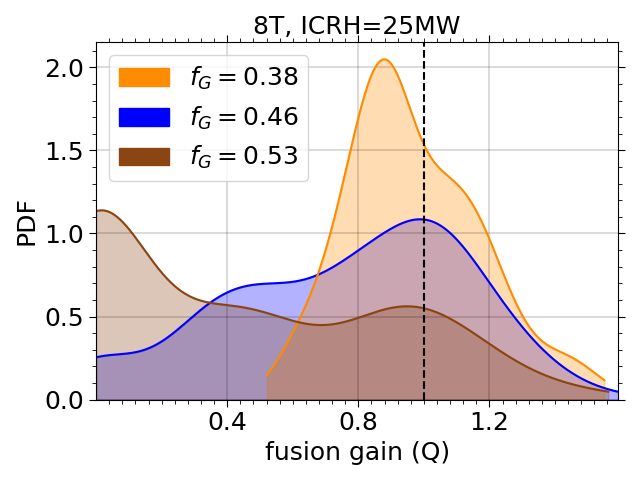}
    \includegraphics[width=0.45\textwidth]{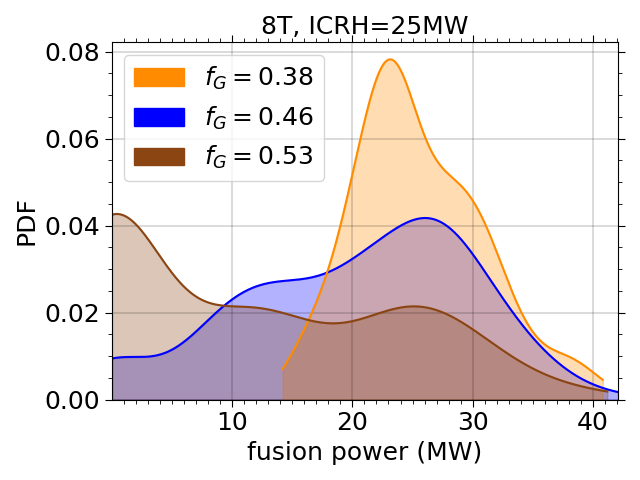}
    \includegraphics[width=0.45\textwidth]{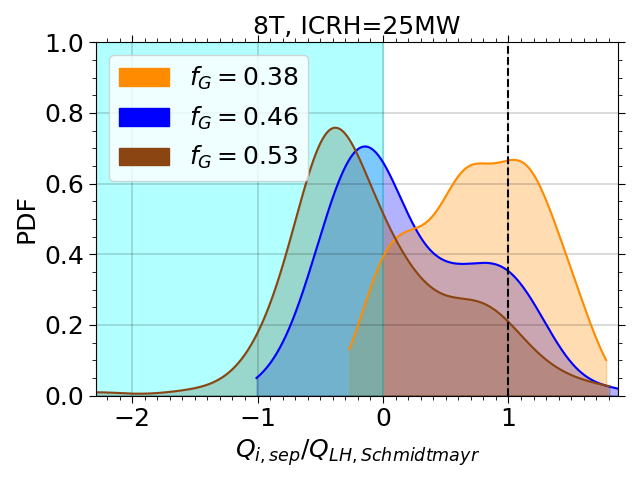}
    \includegraphics[width=0.45\textwidth]{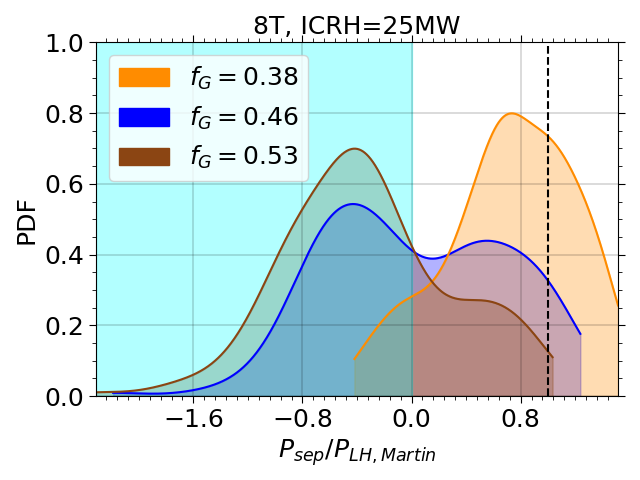}
    \caption{From the top left, clockwise: PDF of the fusion gain, fusion power, $f_{LH}$ with the Martin and Schmidtmayr scalings. An average $Q\sim 1$ with low fusion power (i.e. low neutron load on magnetic coils) has been found for the lowest density, while the H-mode is poorly sustained according to the mentioned regressions, especially for higher densities. A slight difference is found for $f_{LH}$, using ion and total power heat flux scalings.}
    \label{fig:8T_PDFs}
\end{figure}

\begin{figure}[h]
    \centering
    \includegraphics[width=0.45\textwidth]{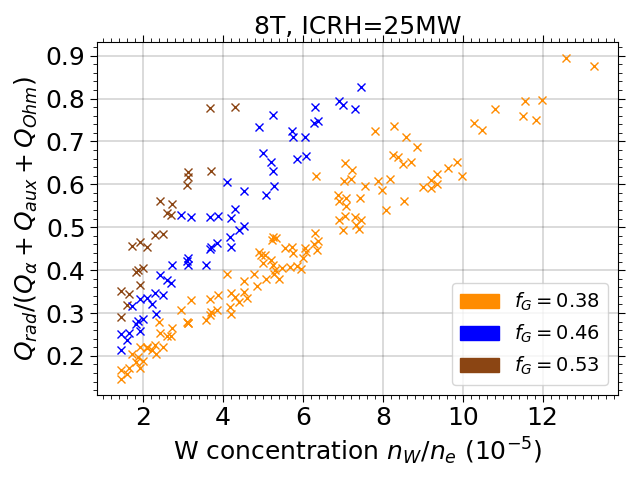}
    \includegraphics[width=0.45\textwidth]{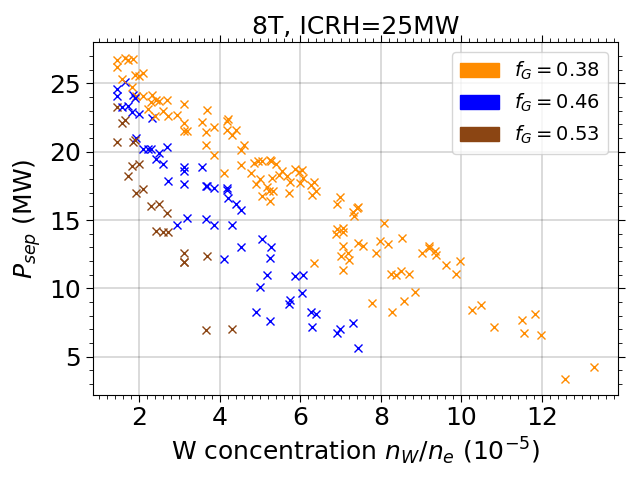}
    \includegraphics[width=0.45\textwidth]{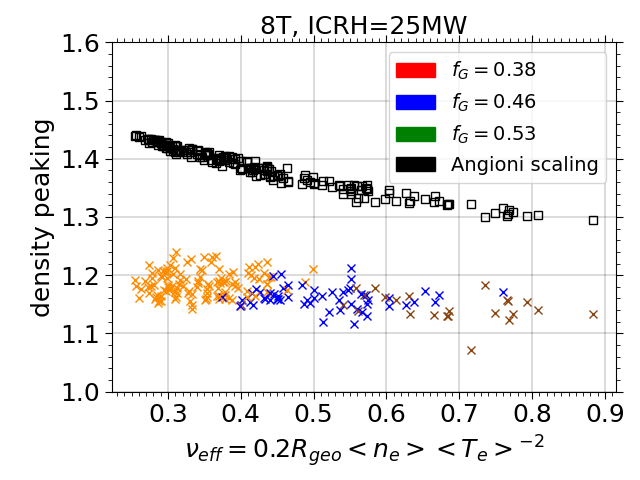}
    \includegraphics[width=0.45\textwidth]{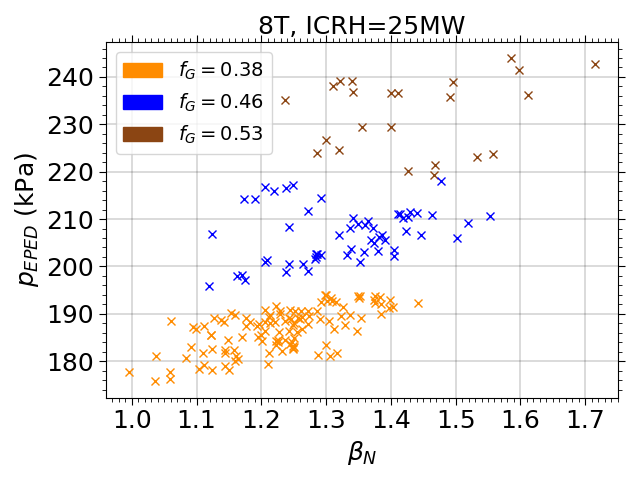}
    \caption{From top left, clockwise: $Q_{rad}/Q_{input}$ and $P_{sep}$ vs W concentration, $p_{EPED}$ vs $\beta_N$, density peaking and the expected value by Angioni scaling \cite{angioni_scaling_2007} vs effective collisionality. The upper plots show that at higher density the radiation fraction increases and the power at the separatrix diminishes, meaning that the H-mode is less sustained. The bottom plots show that increasing the density (i.e. $\beta_N$ and $\nu_{eff}$) the pedestal is higher and the density peaking is nearly insensitive, implying that at higher density a higher fusion power is expected. The cyan region indicates unconverged simulations.}
    \label{fig:8T_Qrad_nu}
\end{figure}

Starting from the initial database described in the previous section, a scan in pedestal density has been performed, to study the effect on fusion performance and H-mode sustainment. The results are shown in figure \ref{fig:8T_PDFs}, through probability distribution functions of $Q$, $P_{fus}$ and $f_{LH}$, calculated according to both Schmidtmayr and Martin scaling.
It is observed that decreasing the density leads to lower fusion power, which, with constant total input power, results in lower fusion gain. This behaviour was found also for the full-field H-mode case. However, at lower density the H-mode is better sustained, as can be seen in the bottom plots of the figure. This is in part due to the reduction of the LH power threshold, as previously discussed. Additionally, in this specific scenario, lower density results in increased $P_{sep}$ due to the simultaneous reduction in $Q_{rad}$ and $Q_{rad}/Q_{input}$. This observation is supported by the fact that increasing the density reduces $P_{sep}$, showing many negative values, which correspond to unconverged simulations (cyan region of the plot). While the PRD showed a separatrix power insensitive to the density, the lower level of fusion for the reduced-field scenario makes the radiation more significant, increasing the weight of $Q_{rad}$ in the overall power balance (even at low W concentrations). Moreover, the ICRH power, constant with density, is higher than it was for the PRD, reducing the weight of $P_{\alpha}$ in the total input power. $Q_{rad}/Q_{input}$ and $P_{sep}$ are shown in figure \ref{fig:8T_Qrad_nu}, along with the density peaking and the top pedestal pressure predicted by EPED-NN. The latter 2 variables show that increasing density leads to higher fusion power, thanks to higher pedestal pressure and overall improved core profile. The same result was found for the PRD.
Figure \ref{fig:8T_PDFs} shows that the mean fusion gain is approximately 1 for lower densities, making this scenario breakeven relevant. Interestingly, the Schmidtmayr scaling shows slightly more optimistic H-mode sustainment with respect to Martin scaling, suggesting that the equipartition between ion and electron heat fluxes plays an important role for the 8T H-mode. This small discrepancy may be connected to the collisional exchange between ions and electrons, which change strongly with $T_i/T_e$, and the role played by radiation, which impacts only the electron heat flux. Since radiation is particularly detrimental for the 8T H-mode, it has a strong impact on the $Q_e$ at the separatrix, leading to an overall higher $Q_i/Q_e$ ratio, and in turn to higher $f_{LH,i}/f_{LH}$, where $f_{LH,i}=P_{sep,i}/P_{LH,Schmidtmayr}$ and $f_{LH}=P_{sep}/P_{LH,Martin}$. However, no deeper investigation has been conducted on this discrepancy, due to model limitations in accurately reproducing SOL, edge, and neutral physics, which would be essential for a precise description of the LH transition.
\begin{figure}[t]
    \centering
    \includegraphics[width=0.45\textwidth]{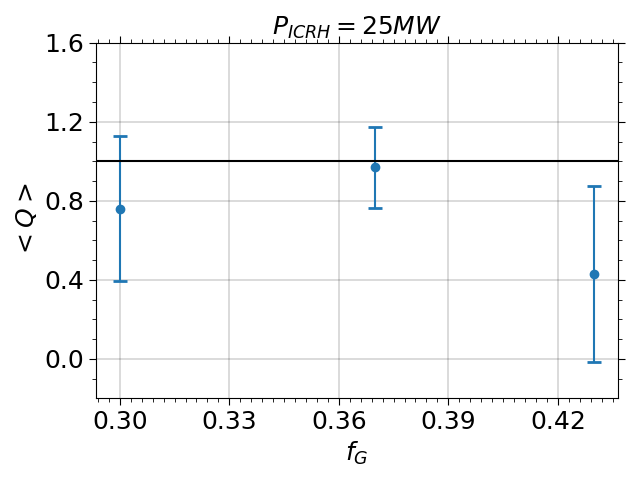}
    \includegraphics[width=0.45\textwidth]{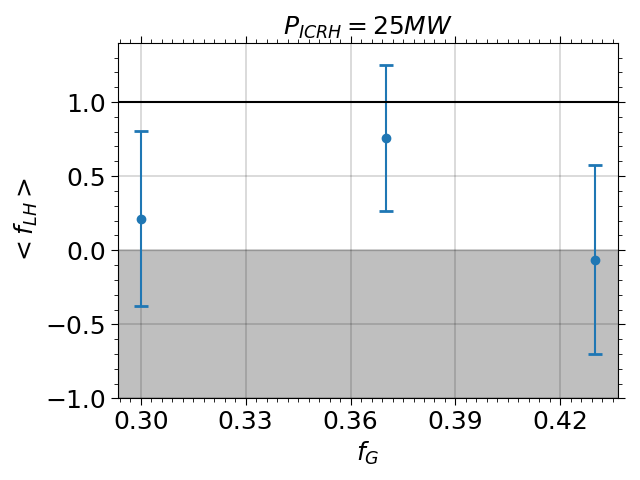}
    \caption{On the left (right) the mean values and standard deviations of $Q$ ($f_{LH}$, according to Martin scaling) for different Greenwald fractions. The gray region indicates unconverged simulations.}
    \label{fig:8T_mean_std}
\end{figure}
\begin{figure}[h]
    \centering
    \includegraphics[width=0.4\textwidth]{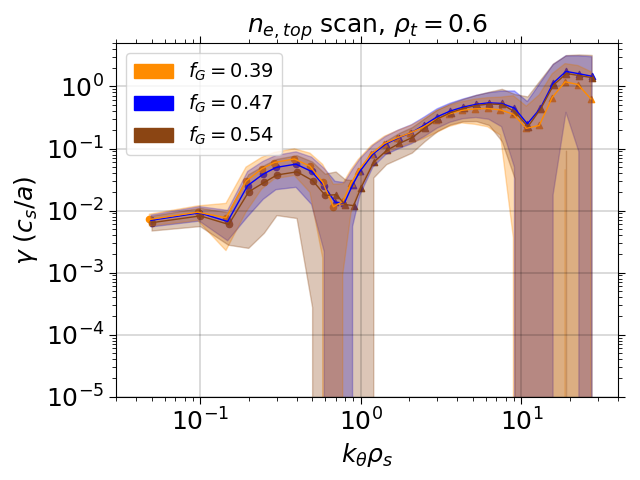}
    \includegraphics[width=0.4\textwidth]{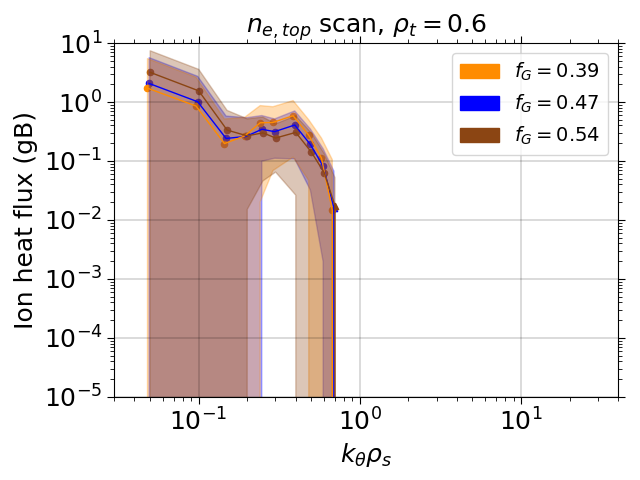}
    \includegraphics[width=0.4\textwidth]{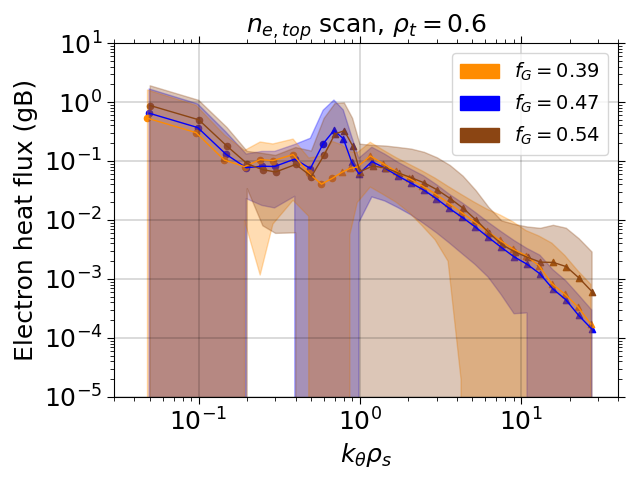}
    \caption{At the top left (right) the mean and standard deviation of TGLF growth rates (ion heat fluxes) at $\rho_t=0.6$, for a pedestal density scan around the reference operational point found for the SPARC 8T H-mode. At the bottom the mean and standard deviations for the electron heat flux. The spectra are very similar across the scan, showing a weakly higher electron flux at high-k scale for higher densities. The big lower error bars are due to stable modes in the TGLF simulations, and are emphasized by the logarithmic scale.}
    \label{fig:8T_spectra}
\end{figure}
To summarize the results found for the SPARC 8T H-mode, the mean values and standard deviations of $Q$ and $f_{LH}$, calculated with the Martin scaling, are plotted in figure \ref{fig:8T_mean_std} for each $f_G$. While $Q\sim 1$ is satisfied in the upper half part of the error bar for low/intermediate densities, the average H-mode is not well sustained, due to the important role played by radiation at high W concentrations.

\subsection{TGLF spectrum analysis}
\begin{figure}[h]
    \centering
    \includegraphics[width=0.45\textwidth]{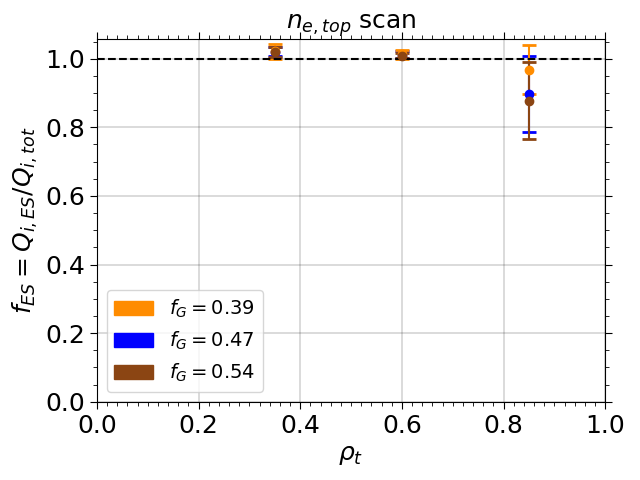}
    \includegraphics[width=0.45\textwidth]{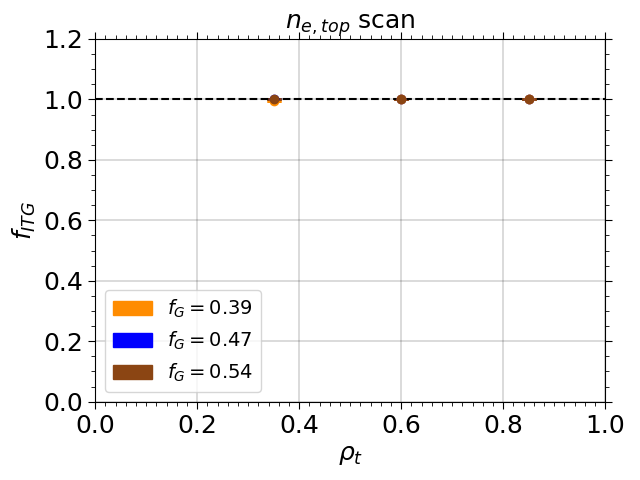}
    \includegraphics[width=0.45\textwidth]{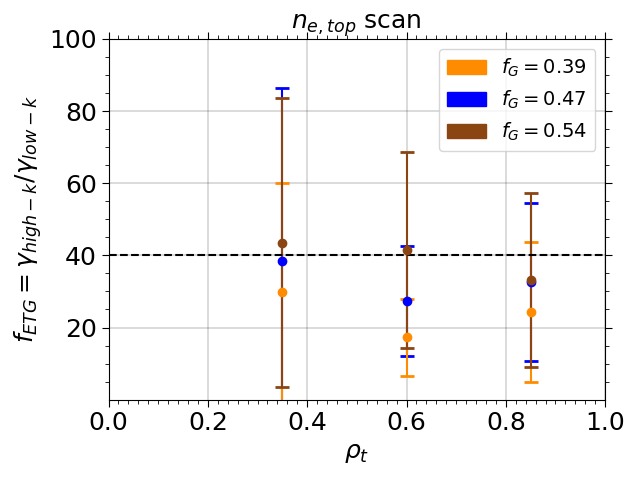}
    \includegraphics[width=0.45\textwidth]{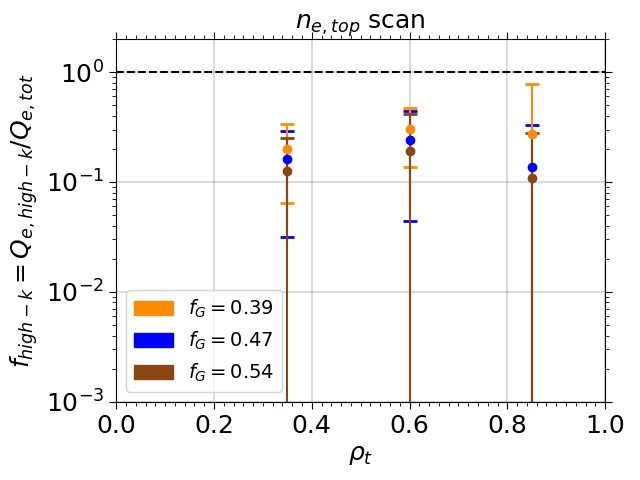}
    \caption{From the top left, clockwise: the average value and standard deviation of $f_{ES}$, $f_{ITG}$, $f_{high-k}$ and $f_{ETG}$ at different radial locations for a density scan. The scan shows robust electrostatic ion heat flux, ITG dominance at low-k scales, increased ETG linear activity at higher density and low electron heat flux at high-k scales.}
    \label{fig:8T_spectra_global}
\end{figure}

As for the PRD, the TGLF spectra have been analyzed for the 8T H-mode. The spectrum of the growth rate, ion and electron heat flux for $\rho_t=0.6$ is shown in figure \ref{fig:8T_spectra}. The spectrum of $\gamma$ and $Q_i$ is very similar for the different densities, showing ITG dominance at low-k. This dominance is confirmed by linear proportional trends of $Q_i$ and $Q_e$ with $a/L_{T_i}$ and by the $f_{ITG}$ value at different radial locations, which is shown in the upper right plot of figure \ref{fig:8T_spectra_global}. It is also consistent with $Q_i/Q_e>1$ ratios found in the simulations at most radial locations. In figure \ref{fig:8T_spectra_global}, it is also shown that the ion heat flux has been found often purely electrostatic along the radius. The general turbulence features found for this scenario are the same described for the PRD. Moreover, a small increase of electron heat flux at very high-k electronic scales is found at higher density, suggesting an associated moderate ETG activity. This is confirmed by $f_{ETG}$ in figure \ref{fig:8T_spectra_global}, which increases at higher density up to an average value of around 40. However, as found for the PRD, the increment in the ETG activity does not couple to significant increment in the fraction of electron heat flux arising at high-k.

\section{Summary and conclusions}
Two SPARC scenarios have been studied in this article: the full field (PRD, 12.2T) and the reduced field (8T) H-modes, employing TGLF for the core transport and a NN trained on EPED for the pedestal stability. To assess the variability of fusion performance 4 input parameters, $f_W$, $f_{DT}$, $T_{i,top}/T_{e,top}$ and $p_{top}/p_{EPED}$, were assigned with random uniform distributions within ranges reflective of those typically observed in experiments. The sensitivity studies revealed that the pedestal pressure and ion to electron temperature ratio are the primary parameters affecting the fusion gain, exhibiting linear trends due to the increased fusion power at higher pressure and ion temperature. $f_W$ is also an important parameter, as above a certain W concentration, plasma radiation can not be sustained, leading to a plasma collapse. Furthermore, increasing radiation makes it less likely for the plasma to maintain an H-mode. This motivated a scan of input ICRH power for the PRD, which at the highest $P_{ICRH}$ showed $Q>2$ and sustained H-mode for most of the input assumptions, including very pessimistic combinations. A pedestal density scan was performed to assess the average dependence of fusion gain and $f_{LH}$ on the density. An increase in $Q$ with $n_{e,top}$ was found, while the LH fraction showed minimal variation. The TGLF turbulent spectrum was analyzed, showing robust ITG dominance at low-k, and low ETG activity at high-k scale, which increased with higher density. However, the increased high-k turbulence did not correspond to a significant increase in the fraction of electron heat flux at high scales. \\
The reduced field H-mode operational scenario was explored to find a combination of $f_G$ and $P_{ICRH}$ which optimizes the fusion gain, while simultaneously achieving $Q>1$ and $f_{LH}>1$. This resulted in an operational point of $f_G=0.46$ and $P_{ICRH}=25MW$. A sensitivity study of the fusion performance, with random variations in the input parameters, showed similar features to the PRD case. Subsequently, a density scan showed similar average fusion gain values and more robust H-mode at lower densities. The overall values of $Q$ indicated that the reduced-field H-mode is a promising candidate for the $Q>1$ goal of the early SPARC operations, particularly given its lower plasma energy content, which implies better exhaust handling, reduced technological risks and lower impact on PFCs, such as in the event of a disruption. The ion and total heat flux scalings for the LH transition showed some disagreement, suggesting that the ion-electron heat flux partition and radiation play roles in determining the H-mode access. However, the H-mode sustainment remains an open problem for this scenario, as many plasma conditions yielded $f_{LH}$ values below 1. Nevertheless, it should be noted that the existing scalings for LH access are based on experimental data from current devices with wide error bars. While these scalings are useful for present devices, their reliability in predicting H-mode access for future devices is uncertain. The results found for the 8T H-mode suggest that the "low $P_{ICRH}$ - low $f_G$" combination can sustain H-mode while maintaining $Q>1$. However, the proportional relation between $P_{LH}$ and average density reverses its trend at low $\langle n_e\rangle$ \cite{hughes_projections_2020}, making this region of the operational space inaccessible. Therefore, future efforts for this scenario should focus on a more detailed study of LH access. As it was for the full-field scenario, TGLF spectra showed ITG dominance and an increased high-k activity at higher densities, not associated with a significant increase in electron heat flux.\\
This work has shown that input assumptions in transport simulations can have a strong impact on confinement and fusion predictions, even when these assumptions lie within realistic ranges and error bars based on existing models and experimental data. Therefore, it is crucial to quantify the variability of fusion-relevant parameters, such as fusion power and gain, in relation to these assumptions, especially for future reactors. Future work may extend the sensitivity study to different scenarios and reactor-relevant machines, incorporate other relevant assumptions as free parameters, and improve the models included in the framework (e.g., self-consistent separatrix evolution, simplified SOL models).

\section*{Acknowledgments}
The authors thank the MIT PSFC for interesting scientific discussions, in particular the MFE Integrated Modeling group, John Wright for technical support about computational resources, Devon Battaglia for insights on the SPARC scenarios physics, Jerry Hughes and Amanda Hubbard for the edge physics support and knowledge offered, Teobaldo Luda, Daniel Fajardo and Michael Bergmann for sharing experiences about code usage. The authors acknowledge the use of ChatGPT during the article editing phase. This research used resources of the National Energy Research Scientific Computing Center, a DOE Office of Science User Facility using NERSC award FES-ERCAP0032161, for the EPED simulations used to train the neural nework model. The ASTRA simulations (ASTRA from main branch with hash d609be34d142987340150d107879d3526dfd5864) presented in this paper were performed on the MIT-PSFC partition of the Engaging cluster at the MGHPCC facility (www.mghpcc.org) which was funded by DoE grant number DE-FG02-91-ER54109. \\
\textit{This work was supported by CFS under RPP020 fundings}.\\

\bibliographystyle{iopart-num} % Use the IOP style file
\bibliography{references} % The name of your .bib file (omit the .bib extension)

\end{document}